\documentclass[oneside,11pt]{article}

\usepackage[utf8]{inputenc}
\usepackage[T1]{fontenc}

\usepackage{amsthm}
\usepackage{amsmath}
\usepackage{amsfonts}
\usepackage{amssymb}

\usepackage{graphicx} 
\usepackage{float}
\usepackage{booktabs}
\usepackage{multirow}
\usepackage{rotating}
\usepackage{array}
\usepackage{xcolor}
\usepackage{subfig}
\usepackage[ruled]{algorithm2e}

\newcolumntype{R}[1]{>{\raggedleft\let\newline\\\arraybackslash\hspace{0pt}}m{#1}}
\newcolumntype{L}[1]{>{\raggedright\let\newline\\\arraybackslash\hspace{0pt}}m{#1}}

\usepackage{breakcites}

\usepackage{nowidow}

\usepackage{enumitem}

\usepackage[top=1.5cm,bottom=2cm,right=2.5cm,left=2.5cm]{geometry}
\usepackage{titling}

\usepackage{natbib}

\usepackage{ifplatform}

\usepackage{textcomp}

\ifwindows
\fi

\allowdisplaybreaks

\DeclareFontFamily{OT1}{pzc}{}
\DeclareFontShape{OT1}{pzc}{m}{it}{<-> s * [1.10] pzcmi7t}{}
\DeclareMathAlphabet{\mathpzc}{OT1}{pzc}{m}{it}

\newtheorem{theorem}{Theorem}

\usepackage[pdftex,final,bookmarksnumbered,bookmarksopen=false,breaklinks,colorlinks]{hyperref}
\hypersetup{
	final,
	bookmarksnumbered=true,
	bookmarksopen=true,
	bookmarksopenlevel=0,
	unicode=false,          % non-Latin characters in Acrobat?s bookmarks
	pdftoolbar=true,        % show Acrobat?s toolbar?
	pdfmenubar=true,        % show Acrobat?s menu?
	pdffitwindow=false,     % window fit to page when opened
	pdftitle={Scaled torus principal component analysis},    % title
	pdfdisplaydoctitle=true, %display document title instead of file name in title bar
	pdftoolbar=true,
	pdfmenubar=true,
	pdflang={English},
	pdfauthor={Pavlos Zoubouloglou, Eduardo Garcia-Portugues, J. S. Marron},     % author
	pdfsubject={arXiv paper},   % subject of the document
	pdfcreator={Eduardo Garcia-Portugues},   % creator of the document
	pdfproducer={Eduardo Garcia-Portugues}, % producer of the document
	pdfkeywords={Dimension reduction}{Directional statistics}{Multidimensional scaling}{Principal component analysis}{Statistics on manifolds}, % list of keywords
	pdfnewwindow=true,      % links in new window
	breaklinks=true,
	hidelinks,       
	linkcolor=black,          % color of internal links (change box color with linkbordercolor)
	citecolor=black,        % color of links to bibliography
	filecolor=magenta,      % color of file links
	urlcolor=cyan           % color of external links
}

\newif\ifmain
\maintrue
\newif\ifsupplement
\supplementtrue

\newif\iffigstabs
\figstabstrue

\usepackage[symbol]{footmisc}
\makeatletter
\newcommand{\ssymbol}[1]{^{\@fnsymbol{#1}}}
\makeatother

\usepackage{bm}

\begin{document}

\ifmain

%-----------------------------------------------%
\title{Scaled torus principal component analysis}
\setlength{\droptitle}{-1cm}
\predate{}%
\postdate{}%
\date{}
%-----------------------------------------------%

%-----------------------------------------------%
\author{Pavlos Zoubouloglou$^{1,3}$, Eduardo Garc\'ia-Portugu\'es$^{2}$, and J. S. Marron$^{1}$}
\footnotetext[1]{Department of Statistics and Operations Research, University of North Carolina at Chapel Hill (USA).}
\footnotetext[2]{Department of Statistics, Carlos III University of Madrid (Spain).}
\footnotetext[3]{Corresponding author. e-mail: \href{mailto:pavlos@live.unc.edu}{pavlos@live.unc.edu}.}
\maketitle
%-----------------------------------------------%

\begin{abstract}
	A particularly challenging context for dimensionality reduction is multivariate circular data, i.e., data supported on a torus. Such kind of data appears, e.g., in the analysis of various phenomena in ecology and astronomy, as well as in molecular structures. This paper introduces Scaled Torus Principal Component Analysis (ST-PCA), a novel approach to perform dimensionality reduction with toroidal data. ST-PCA finds a data-driven map from a torus to a sphere of the same dimension and a certain radius. The map is constructed with multidimensional scaling to minimize the discrepancy between pairwise geodesic distances in both spaces. ST-PCA then resorts to principal nested spheres to obtain a nested sequence of subspheres that best fits the data, which can afterwards be inverted back to the torus. Numerical experiments illustrate how ST-PCA can be used to achieve meaningful dimensionality reduction on low-dimensional torii, particularly with the purpose of clusters separation, while two data applications in astronomy (three-dimensional torus) and molecular biology (on a seven-dimensional torus) show that ST-PCA outperforms existing methods for the investigated datasets.
\end{abstract}
\begin{flushleft}
\end{flushleft}

%---------------------------%
\section{Introduction}
\label{sec:intro}
%---------------------------%

% Introduction
Recent advances in the way information is collected have brought along the need to analyze data of a non-Euclidean nature. Important among these are multivariate circular data, or toroidal data, which are vectors of angles that are naturally represented on the torus $\mathbb{T}^d := [-\pi, \pi)^d$, $d\geq1$, where $-\pi$ and $\pi$ are identified. An important instance of toroidal data happens in molecular biology, where vectors of torsion angles characterize protein and RNA structures \citep{Ramachandran1963,Duarte1998}. A different source for data on $\mathbb{T}^d$ is environmental sciences, with planar directions being key constituents of wind and sea currents vector fields; see \cite{Ducharme2012} for an application in the latter context. Another popular instance of non-Euclidean data is spherical data, i.e., data on $\mathbb{S}^d:=\{\bm{x}\in\mathbb{R}^{d+1}:\|\bm{x}\|=1\}$ with $d = 2$. Among other fields, spherical data plays a relevant role in astronomy; see \cite{Jupp2003} for a statistical analysis of the normal vectors associated to long-period comet orbits. \cite{directional-survey} and \cite{MarronOODA} provide recent surveys of existing statistical methodology for data on spheres and torii, and for other types of non-Euclidean spaces, respectively.\\

% PCA in manifolds -- early developments
Principal Component Analysis (PCA) is arguably the most popular multivariate technique, its adaptation to non-Euclidean data being the subject of considerable research in the last decades. A first step towards identifying modes of variation for manifold-valued data were the tangent plane-based methods, which seek to utilize the locally Euclidean property of manifolds. An important representative of them is the Principal Geodesic Analysis (PGA) of \cite{PGA}. Recall that (classic) Euclidean PCA builds the directions of maximal projected variation starting at the sample mean. PGA parallels this by starting with the Fr\'echet (sample) mean and finding the geodesic through that mean on which the projected variation in the tangent plane is maximal. Thus the first principal geodesic is obtained and the remaining modes of variation are obtained sequentially, by imposing orthogonality on geodesics at the Fr\'echet mean. \cite{Geodesic-PCA} showed that the constraint of geodesics passing through that mean seriously limited the flexibility of PGA and proposed Geodesic-PCA, which removes precisely that constraint. Later, \cite{Huckemann2010} proposed generalized geodesics and orthogonal projections on quotient spaces, and used these as components for a PCA on certain Riemannian manifolds.\\

% PNS and PCA in manifolds
In the context of skeletal models for object shape representations, modes of variation on the sphere that follow non-great subspheres (i.e., non-geodesics) can provide more effective dimensionality reduction, as noted, e.g., in \cite{pizer2017object} and \cite{pizer2020}. This motivated the principal arc analysis of \cite{Principal-Arc-Analysis} and its generalization on $\mathbb{S}^d$, $d\geq2$, \cite{PNS}'s Principal Nested Spheres (PNS). PNS exploits the fact that intersecting $\mathbb{S}^d$ with a hyperplane of its Euclidean embedding yields a subsphere isomorphic to $\mathbb{S}^{d-1}$ which is not necessarily a great subsphere. Therefore, PNS offers a richer family of possible modes of variation on $\mathbb{S}^d$ than manifold-generic methods for dimensionality reduction, that are focused on great subspheres only. This provided evidence for the utility of exploiting the specificity of the data support in building a more informative dimension-reduction analysis. PNS features an iterative process that yields a sequence of best-fitting subspheres of decreasing dimensions to $\mathbb{S}^d$. This motivated the concept of ``backwards analogues'' of PCA, as discussed in \cite{Backwards-PCA}: for a $d$-dimensional manifold, a $(d-1)$-dimensional best-fitting submanifold is identified, and then this process is iterated, in a nested fashion, until a one-dimensional mode of variation has been obtained. Under fairly general assumptions, \cite{huckemann2018} proved that backwards component sequences (including PNS) satisfy asymptotic results, such as strong consistency and asymptotic normality. Yet both forward and backward analogues of PCA are greedy algorithms, which do not necessarily guarantee a global optimum in the sequence of proportion of explained variance. In response to this point, \cite{pennec2018} developed barycentric subspace analysis, which introduces a nested (backwards or forward) sequence of submanifolds that is defined as the locus of points that are manifold-affine spans of a set of $d+1$ points in a $d$-dimensional manifold (great subspheres for $\mathbb{S}^d$).\\

% Torus PCA I: Euclidean arrival space
Despite its apparent simplicity, the torus is a particularly challenging manifold for dimensionality reduction. Tangent plane methods disregard long-range periodicity and tend to underperform, unless the data are fairly closely distributed. An attempt to adapt geodesic-based methods on the torus is torus principal geodesic analysis by \cite{Nodehi2015}. Yet, as noted in \cite{Eltzner2018}, geodesics on the torus face the problem of producing space-filling curves almost surely by winding around indefinitely. For that reason, existing methods on PCA adaptations for toroidal data can be classified into those relying either on (a) imposing a certain distributional model on the data or (b) transforming the data to a space where dimension-reduction tools are already consolidated. With respect to approach (a), one can find \cite{Kent2009}'s approach of PCA on the covariance matrix of a wrapped normal model on the torus. A recent approach proposed by \cite{nodehi2020torus} extends probabilistic PCA to Torus Probabilistic PCA (TPPCA), where the toroidal data are expressed as a latent variable model modulo $2 \pi$, while assuming Gaussianity for the latent variables. The parameters of the model are then to be estimated by an Expectation--Maximization algorithm. With regard to approach (b), typically the arrival space has been set as the Euclidean space or variations thereof. For example, \cite{Mu2005} proposed dihedral PCA (dPCA), which leverages the usual embedding $\mathbb{T}^d \rightarrow [-1,1]^{2d}$ via sines and cosines, and then performs PCA on $\mathbb{R}^{2d}$, a drawback being that ordinary PCA ignores the fact that the data lie on a curved $d$-dimensional manifold within $[-1,1]^{2d}$. A similar transformation is that of \cite{Altis2007}, using the complex representation of the angles. \cite{Riccardi2009} propose angular PCA (aPCA), where they first center the toroidal data and then proceed with usual PCA. \cite{Sittel2017} modified aPCA to dPCA+, a variant that attempts to minimize the distortion by shifting the ``boundaries'' of $\mathbb{T}^d$\nopagebreak[4] to the lowest density region.\\

% Torus PCA II: sphere arrival space
Still within (b), but differently to the previous approaches, \cite{Sargsyan2012}'s Geodesic PCA deforms $\mathbb{T}^d$ to $\mathbb{S}^d$, then performs dimensionality reduction in the latter space by PGA. Nevertheless, this transformation directly uses the angles in $\mathbb{T}^d$ as the $d$ angles that determine the hyperspherical coordinates of $\mathbb{S}^d$, with the consequent appearance of singularities (e.g., when $d=2$ the two maximally-separated points $(\pm\pi/2,\pm\pi/2)$ on $\mathbb{T}^2$ are mapped to the same point on $\mathbb{S}^2$). Perhaps the most successful and advanced existing method for PCA on the torus is \cite{Eltzner2018}'s Torus PCA (T-PCA), where $\mathbb{T}^d$ is also deformed to $\mathbb{S}^d$, now with a more sophisticated transformation, and PNS is then applied. However, the deformation in T-PCA is not invariant to permutations of the angular variables, which introduce a source of subjectivity in the analysis. To alleviate this issue, the authors propose two data-driven orderings of variables, \emph{SI ordering} and \emph{SO ordering}, corresponding to sorting the variables in terms of descending and increasing amount of circular spread, respectively. Yet the practitioner has to choose one ordering, the two possibly yielding significantly different scores that may affect subsequent analyses (as evidenced in Figure \ref{fig:RNA}). Furthermore, the deformation to $\mathbb{S}^d$, although designed to cut open the torus at a point with the minimum distortion, may induce relevant artifacts to certain datasets. These deformation artifacts can create spurious cluster structures in the scores of datasets with non-existent cluster structures (see Figure \ref{fig:sunspots_hist}). A final limitation of T-PCA, presented in Remark 2.3 of \cite{Eltzner2018}, is that T-PCA is applicable whenever there is a structural data gap in all angles except for at most two.

\begin{figure}[!h]
	\centering
	\subfloat[]{\includegraphics[width=0.3\textwidth]{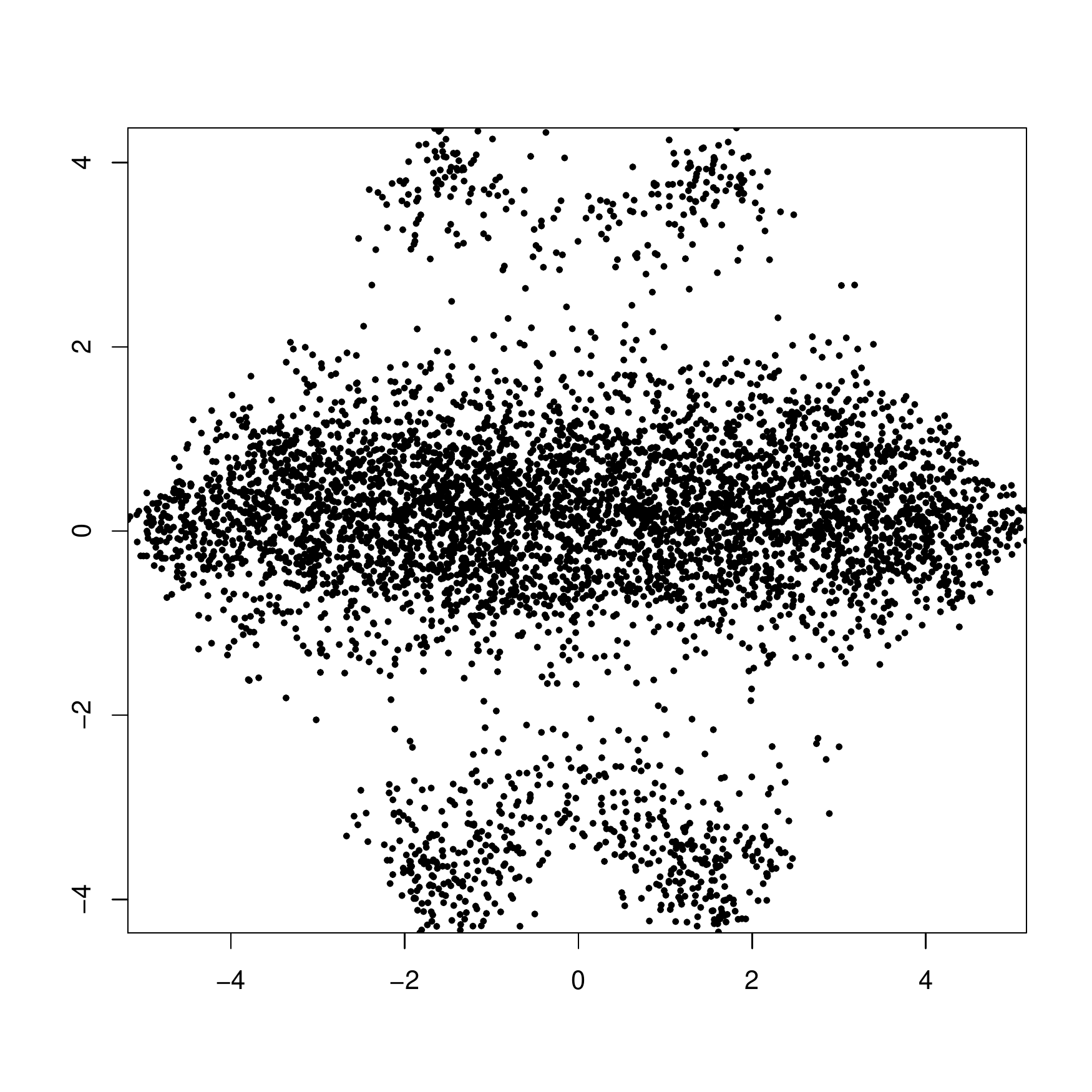}\label{fig:pns_E}}
	\hfill
	\subfloat[]{\includegraphics[width=0.3\textwidth]{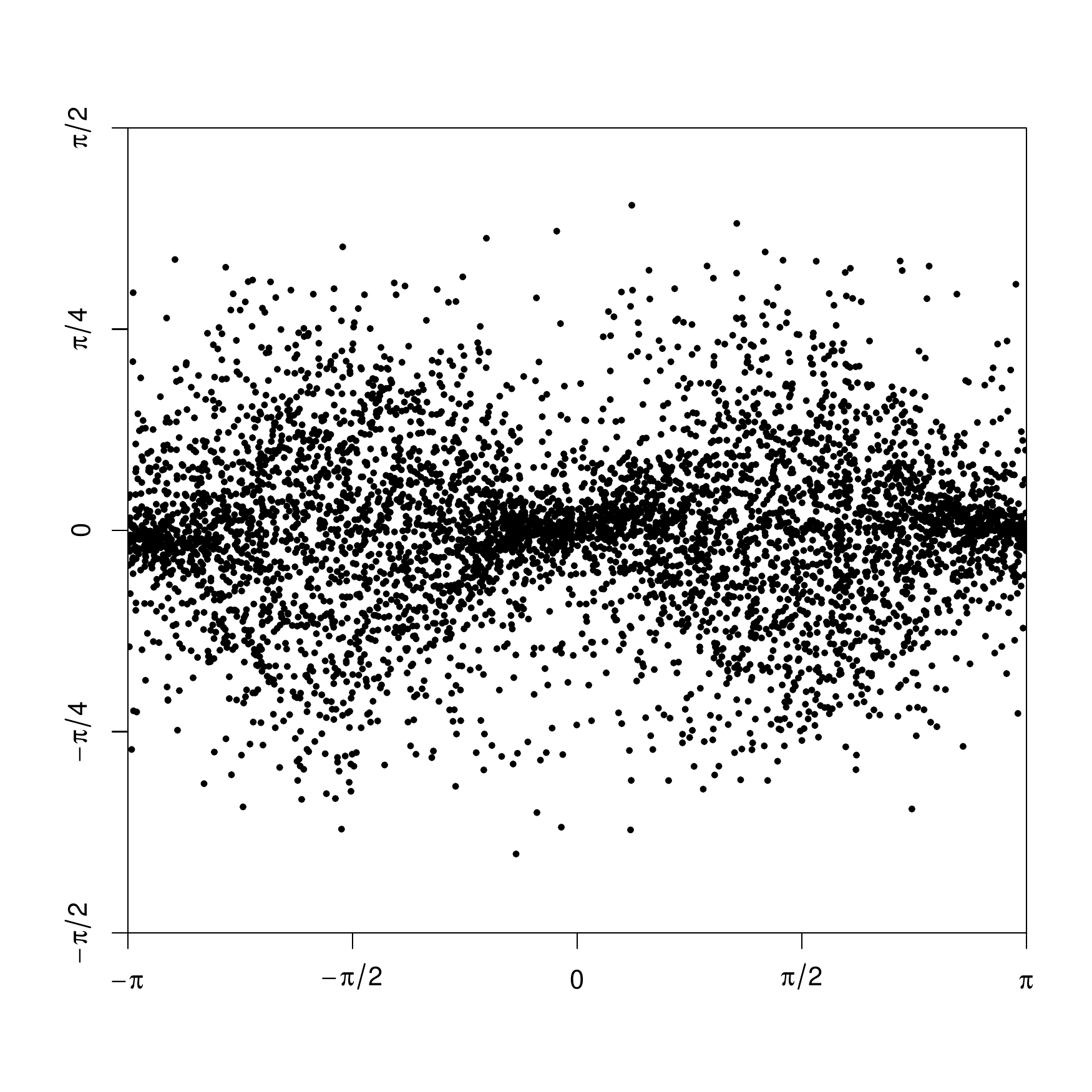} \label{fig:ST-PCA_E}}
	\hfill
	\subfloat[]{\includegraphics[width=0.37\textwidth]{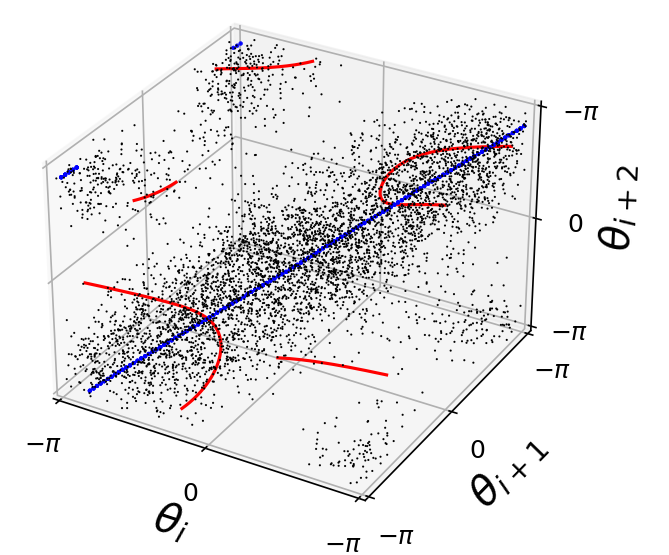} \label{fig:STPCA-TPCA_Variation}}
	\caption{\small Scores (scatterplots (a) and (b)) and modes of variation (panel (c)) for the sunspots data; panels (a) and (b) show the first (horizontal axis) and second (vertical axis) scores for PCA and ST-PCA, respectively. In panel (c), the original data is shown with black points, while the T-PCA and ST-PCA first modes of variation are represented as red and blue solid lines, respectively. The figure illustrates that ST-PCA gives a more efficient mode of variation than standard PCA and T-PCA.}
	\label{fig:Sunspots_E_Variation}
\end{figure}

% Our contributions
In this paper we propose a toroidal PCA that is based on embedding the sample on $\mathbb{T}^d$ into $\mathbb{S}^d$ by using a data-driven map aimed at preserving the pairwise geodesic distances of the data. Our dimension-reduction method is summarized in three steps. First, we transform the original sample on $\mathbb{T}^d$ to a similar configuration of points on a $d$-dimensional sphere by using Spherical MultiDimensional Scaling (SMDS). SMDS is an optimization procedure designed to minimize the squared differences between the sample pairwise geodesic distances on $\mathbb{T}^d$ and the corresponding pairwise geodesic distances of the transformed sample on a $d$-sphere. The radius of the arrival $d$-sphere is also chosen automatically in a further attempt to minimize distortions. Second, we apply PNS to the transformed data to find the sequence of best-fitting subspheres and the scores associated to them. Finally, we show how to optionally invert the SMDS map, in a step usually referred to as \emph{prediction}, to produce a low-dimensional representation of the data in the original torus via the inverted PNS scores. Due to the multidimensional scaling operation, we henceforth refer to our method as Scaled Torus PCA (ST-PCA). An illustration of the adequacy of ST-PCA for data analysis is advanced in Figure \ref{fig:Sunspots_E_Variation}. Figure \ref{fig:STPCA-TPCA_Variation} shows astronomical data on the appearance of sunspots, which lie on $\mathbb{T}^3$ in a roughly diagonal pattern, and are later described in detail in Section \ref{subsec:Sunspots}. Figure \ref{fig:pns_E} displays the scores from standard PCA, which disregards the periodicity of the data, while Figure \ref{fig:ST-PCA_E} shows that ST-PCA gives more sensible scores. Figure \ref{fig:STPCA-TPCA_Variation} shows that while T-PCA honors the periodicity of the data, it yields a less efficient mode of variation (red curve) when compared to ST-PCA (blue curve). In this case study, and among other competitors, ST-PCA performs the best in the sense of preserving data fidelity and explaining the most variance in the first principal component. A second case study to an RNA dataset in $\mathbb{T}^7$ that has lately been used as benchmark in the literature also evidences the adequacy of ST-PCA among other existing alternatives, in the sense of correctly clustering the largest amount of points based on the first principal component scores. We evidence also the ability of ST-PCA to perform dimension reduction and identifying clusters in the data with numerical experiments on different data patterns on $\mathbb{T}^2$ and $\mathbb{T}^3$.\\

% Paper organization
The rest of this paper is organized as follows. Section \ref{sec:theory} explores the links between torii and spheres and provides a review of multidimensional scaling, with discussion on relevant literature. In Section \ref{sec:method} we develop ST-PCA in full detail and discuss the subtleties on its implementation. A simulation study on clustered data is provided in Section \ref{sec:simul} to illustrate the strengths of ST-PCA. We apply ST-PCA to two real datasets in Section \ref{sec:real}, evidencing the benefits of ST-PCA with respect to available alternatives. The paper concludes with a discussion in Section \ref{sec:epilogue}.

%---------------------------%
\section{Background}
\label{sec:theory}
%---------------------------%

The first subsection introduces some required notions from geometry and provide justification on why the sphere is a more natural arrival space than the Euclidean space. The reader who is curious about further relevant geometric concepts is referred to \cite{Huckemann2010} for a concise summary that is very accessible to statisticians. The second subsection provides details on classical Euclidean MDS and its spherical variant. An overview of related work is provided.

%---------------------------%
\subsection{Some geometrical considerations}
%---------------------------%

% Data embeddings
Traditional data embeddings into metric spaces take place in low-dimensional Euclidean spaces, as these allow for easiest visualization and inference. Results for isometric embeddings to the Euclidean space are due to the Nash embedding theorem for Riemannian manifolds and \cite{Morgan74} for arbitrary metric spaces. Recent advances in machine learning and computer graphics have popularized the idea of embedding data that lie in inherently curved spaces into lower-dimensional non-Euclidean spaces, as these spaces may reduce the distortion of the embedding.\\

% Basics of S^d and T^d, and notation
In our setting, the $d$-sphere of radius $r$ and the $d$-torus are specially relevant. These spaces are respectively defined as $\mathbb{S}^d_r := \{ \bm{x} \in \mathbb{R}^{d+1} : \|\bm{x}\| = r \}$, where $\|\cdot\|$ is the Euclidean norm, and $\mathbb{T}^d:=\allowbreak(\mathbb{S}^1)^d:=\mathbb{S}^1\times\stackrel{d}{\cdots}\times\mathbb{S}^1$ or, equivalently and as previously introduced, $\mathbb{T}^d := [-\pi, \pi)^{d}$ with $-\pi$ and $\pi$ identified, since $\mathbb{S}^1$ can also be thought of as the quotient space $\mathbb{R} / (2\pi \mathbb{Z})$. For simplicity of notation, we denote  $\mathbb{S}^d_1$ by $\mathbb{S}^d$. Geodesic distances on $\mathbb{S}^d_r$ are given by
\begin{align} 
	\delta_{\mathbb{S}^d_r}(\bm{x},\bm{y})=r \arccos \left(\bm{x}'\bm{y}/r^2 \right),\quad \bm{x},\bm{y}\in\mathbb{S}^d_r. \label{eq:sdist}
\end{align} 
For angular coordinates on $\mathbb{S}^1$, $\delta_{\mathbb{S}^1}(\phi,\psi) = \min \{ |\phi -\psi|, 2\pi - |\phi - \psi| \}$, $\phi,\psi\in[-\pi,\pi)$. The distance on $\mathbb{T}^d$ benefits from the space's product structure: $\delta_{\mathbb{T}^d}(\bm{\phi},\bm{\psi}) := \big(\sum_{i=1}^d \delta_{\mathbb{S}^1}(\phi_i,\psi_i)^2\big)^{1/2}$, $\bm{\phi},\bm{\psi}\in\mathbb{T}^d$.\\

% Why sphere as an arrival space
The construction of our dimensionality reduction tool rests upon the underlying idea of \cite{Eltzner2018} that spherical embeddings are more appropriate than Euclidean ones when considering transformations from $\mathbb{T}^d$. While a definitely arguable claim, we present next a series of reasons supporting it. First, $\mathbb{S}_r^d$ is a space benign enough to isometrically embed an arbitrary distance under certain conditions. Precisely, \cite{Embedding-Metric} provides necessary and sufficient conditions for an isometric embedding of an arbitrary distance to the sphere, as shown in Theorem 3.22 of \cite{Pattern-Recogn}:
\begin{theorem}[\citealt{Embedding-Metric}]
	An $n \times n$ distance matrix $\bm{D} = (\delta_{ij})_{i,j=1}^n$ can be isometrically embedded to a sphere $\mathbb{S}^m_r$ of radius $r>0$ and dimension $m$ if and only if $\delta_{ij} \leq \pi r$, for all $i,j=1,\ldots,n$, and the matrix $\bm{G} = (\cos(\delta_{ij}/r))_{i,j =1}^n$ is positive semidefinite. Then, the smallest $m$ such that $\bm{D}$ embeds to $\mathbb{S}^m_r$ is $m = \mathrm{rank}(\bm{G})-1$, while the solution is undefined for $\mathrm{rank}(\bm{G})=1$.
\end{theorem}
\noindent Second, both $\mathbb{S}_r^d$ and $\mathbb{T}^d$, endowed with their respective metrics, are compact Riemannian manifolds for every $d \in \mathbb{N}$, unlike $\mathbb{R}^d$ with the usual Euclidean distance. Hence the second similarity of the two manifolds (and dissimilarity with the Euclidean space) comes from topology. Third, $\mathbb{S}_r^d$ has a subspace isomorphic to $\mathbb{S}^1$ (recall, e.g., its parametrization in terms of hyperspherical coordinates) that contributes on preserving the $d$-dimensional periodicity in $\mathbb{T}^d$, as opposed to $\mathbb{R}^d$, where periodicity is nonexistent. Fourth, as noted in the context of data analysis at least as early as in \cite{Sphere-MDS}, convex hulls in $\mathbb{S}^d$ and $\mathbb{T}^d$ present definition issues, in contradiction to $\mathbb{R}^d$, where the convex hull of the data provides a natural notion of boundary. A final, yet very important reason, comes from statistics and concerns the ability of PNS to capture cluster structure when compared to PCA. Consider the problem of a dataset with three normal-like clusters on $\mathbb{S}^d$ and $\mathbb{R}^d$. In $\mathbb{S}^d$ it is always possible to capture three clusters by using a one-dimensional periodic mode of variation ($\mathbb{S}^1$), as long as one allows small circles. However, unless the three clusters have colinear centers, in $\mathbb{R}^d$ one can only flawlessly capture up to two clusters by using a straight line. Therefore, an embedding into $\mathbb{S}^d$ followed by PNS allows for a more efficient cluster hunting than one into $\mathbb{R}^d$ and subsequent~PCA. 

%---------------------------%
\subsection{Multidimensional scaling}
%---------------------------%

MultiDimensional Scaling (MDS) is a well-established method with origins that date to the psychometrics literature, as early as the 1950's. Many reviews and books already exist on the topic, see, e.g., \cite{Modern-MDS}, \cite{Cox2008}, or \cite{Modern-Review}. We review the Spherical MDS literature relevant for our work. We will only discuss and work with what is called metric MDS and in particular, its classical and spherical variant.

%---------------------------%
\subsubsection{Classical Euclidean MDS}
%---------------------------%

Let $\bm{x}_1,\ldots,\bm{x}_n$ be a sample of elements in an arbitrary metric space $(X,\delta_X)$, where $\delta_X$ is a distance function on $X$. For a given $p \in \mathbb{N}$, the metric version of Classical MDS seeks to find a configuration $\widehat{\bm{y}}_1,\ldots,\widehat{\bm{y}}_n\in \mathbb{R}^p$ such that
\begin{align}
	(\widehat{\bm{y}}_1,\ldots,\widehat{\bm{y}}_n)=\arg\min_{(\bm{y}_1,\ldots,\bm{y}_n)\in(\mathbb{R}^{p})^n}\frac{1}{n(n-1)}\sum_{i\neq j}(\delta_X(\bm{x}_i,\bm{x}_j)-\|\bm{y}_i-\bm{y}_j\|)^2.\label{eq:mds}
\end{align}
When $(X,\delta_X) = (\mathbb{R}^d, \|\cdot\|)$ for $d \in \mathbb{N}$, $d \geq p$, \eqref{eq:mds} is essentially a standard PCA up to translation (see, e.g., Section 2.2.7 of \cite{Cox2008}), where $\widehat{\bm{y}}_1,\ldots,\widehat{\bm{y}}_n$ are the projections of $\bm{x}_1,\ldots,\bm{x}_n$ to the best-fitting $\mathbb{R}^p$, up to isometry.

%---------------------------%
\subsubsection{Spherical MDS}
\label{subsec:spherical-mds}
%---------------------------%

In the non-classical context, we consider a second metric space $(Y,\delta_Y)$, not necessarily Euclidean, to which we want to map our original data, see, e.g., \cite{Sphere-MDS,Cox2008}. Then, \eqref{eq:mds} is equivalent to finding a configuration $\widehat{\bm{y}}_1,\ldots,\widehat{\bm{y}}_n\in Y$ such that
\begin{align} 
	(\widehat{\bm{y}}_1,\ldots,\widehat{\bm{y}}_n)=\arg\min_{(\bm{y}_1,\ldots,\bm{y}_n)\in Y^n}\frac{1}{n(n-1)}\sum_{i\neq j} (\delta_X(\bm{x}_i,\bm{x}_j)-\delta_Y(\bm{y}_i,\bm{y}_j))^2. \label{eq:gmds}
\end{align}
The objective function in \eqref{eq:mds} and \eqref{eq:gmds} is referred to as the \emph{stress function} for the given method. In the case that $(Y,\delta_Y) = (\mathbb{S}^d_r,\delta_{\mathbb{S}^d_r})$, we refer to \eqref{eq:gmds} as Spherical MDS (SMDS). Typically, to solve the Spherical MDS problem \eqref{eq:smds} where $(X,d_X)$ is arbitrary, one of two strategies have been used. The first represents the data on $\mathbb{S}_r^d$ with hyperspherical coordinates and numerically optimizes in \eqref{eq:gmds} over $Y = [0,\pi]^{d-1} \times [-\pi,\pi)$. Its main advantage is computational convenience, as for the optimization only some box constraints are required. A nuisance of this approach is that computation of Cartesian coordinates is required for computing geodesic distances on $\mathbb{S}_r^d$, for $d \geq 3$; perhaps for this reason, instances of this approach seem to have focused on $d = 1,2$ \citep{Texture-Mapping,Sphere-MDS,Sphere-MDS-2}. The second strategy employs the Euclidean parametrization, as well as either linear or quadratic constraints that will force the new configuration to lie on a sphere. This, however, may yield a somehow demanding optimization problem. Instances of such work include \cite{Early-Sphere-MDS-1}, \cite{Early-Sphere-2}, and \cite{Constrained-MDS-1}. A particularly successful approach in this subcategory that can solve MDS with quadratic constraints by employing majorization techniques is the \textit{primal method} of MDS-Quadratic (MDS-Q), proposed by \cite{MDS-Majorization}. MDS-Q reduces to Spherical MDS when $Y = \mathbb{S}^d$ in \eqref{eq:gmds}. A faster algorithm is also supported by MDS-Q, referred to as the \textit{dual method} of MDS-Q, which yields \textit{approximately} spherical solutions (and so a projection to the sphere must follow).

%---------------------------%
\section{Scaled Torus-PCA}
\label{sec:method}
%---------------------------%

We provide next a high-level overview of Scaled Torus-PCA (ST-PCA). Given a sample $\bm{x}_1,\ldots,\bm{x}_n \in \mathbb{T}^d$, $d \in \mathbb{N}$, ST-PCA proceeds in three main steps:
\begin{enumerate}[label=\arabic{*}., ref=\arabic{*}]
	\item Obtain $\widehat{\bm{y}}_1,\ldots,\widehat{\bm{y}}_n\in \mathbb{S}_r^d$ by solving the SMDS problem arising from \eqref{eq:gmds} with $(X,\delta_X) = (\mathbb{T}^d,\delta_{\mathbb{T}^d})$ and $(Y,\delta_Y) = (\mathbb{S}^d_r, \delta_{\mathbb{S}^d_r})$:\label{step1}
	\begin{align}
		(\widehat{\bm{y}}_1,\ldots,\widehat{\bm{y}}_n)=\arg\min_{(\bm{y}_1,\ldots,\bm{y}_n)\in {(\mathbb{S}^d_r)}^{n}} \frac{1}{n(n-1)} \sum_{i\neq j}(\delta_{\mathbb{T}^d}(\bm{x}_i,\bm{x}_j)-\delta_{\mathbb{S}^d_r}(\bm{y}_i,\bm{y}_j))^2. \label{eq:smds}
	\end{align}
	\item Obtain $S^j$, $j = d-1, \ldots, 1$, the nested sequence of subspaces that are isomorphic to $\mathbb{S}^j$ and that best fit $\widehat{\bm{y}}_1,\ldots,\widehat{\bm{y}}_n$ according to PNS.\label{step2}
	\item Optionally, ``invert'' $S^1$ through a reverse SMDS problem to obtain a principal curve on $\mathbb{T}^d$.\label{step3}
\end{enumerate}
Step \ref{step1} is discussed in Subsection \ref{subsec:implem}, while considerations on choosing the radius of $\mathbb{S}^d_r$ are addressed in Subsection \ref{subsec:radius}. Steps \ref{step2} and \ref{step3} are elaborated in Subsections \ref{subsec:pns}  and \ref{subsec:pred}, respectively.

%---------------------------%
\subsection{Solving the SMDS problem}
\label{subsec:implem}
%---------------------------%

% Our proposed optimization approach
For a given dataset $\bm{x}_1, \ldots, \bm{x}_n \in \mathbb{T}^d$ and radius $r \in \mathbb{R}_+$, in Step \ref{step1} we seek a solution to the optimization problem \eqref{eq:smds}. While this could be achieved by using MDS-Q, as previously described, we have also explored the unconstrained optimization approach
\begin{align}
	(\widehat{\bm{z}}_1,\ldots,\widehat{\bm{z}}_n) = \underset{(\bm{z}_1,\ldots,\bm{z}_n)\in (\mathbb{R}^{d+1})^n}{\arg\min} \frac{1}{n(n-1)}
	\sum_{i \neq j } \left(\delta_{\mathbb{T}^d}(\bm{x}_i,\bm{x}_j) - r\arccos\left( \frac{\bm{z}_i'\bm{z}_j}{ \|\bm{z}_i\|\|\bm{z}_j\|}\right)\right)^2,
	\label{eq:smds2}
\end{align}
whose solution is then projected as $\widehat{\bm{y}}_i:=r\widehat{\bm{z}}_i/\|\widehat{\bm{z}}_i\|\in\mathbb{S}^d_r$, $i=1,\ldots,n$. An advantage of this relatively simple approach over MDS-Q is that it can be further modified straightforwardly, as later discussed in Section \ref{subsec:radius}. More importantly, even though the framework for a \textit{geodesic} MDS-Q exists, existing software only features chordal distances on $\mathbb{S}^d$. When either solving \eqref{eq:smds} or \eqref{eq:smds2} some care is advised: none have unique global minima, as their objective functions are invariant under rotations of point configurations. In \eqref{eq:smds}, this issue can be addressed by fixing $\bm{y}_1$ to, say, $(0,\ldots,0,1)'$ and optimizing over the $(n-1)$ remaining $(\bm{y}_2,\ldots,\bm{y}_n)$. In \eqref{eq:smds2}, one can fix $\bm{z}_1=(0,\ldots,0,s)'$ and optimize on $(s,\bm{z}_2,\ldots,\bm{z}_n)\in\mathbb{R}_+\times(\mathbb{R}^{d+1})^{n-1}$.\\

% Empirical considerations
We ran experiments comparing the MDS-Q approach with the BFGS algorithm as implemented in R's \texttt{optim} function to solve \eqref{eq:smds3} (a slight modification of \eqref{eq:smds2}). MDS-Q was run as implemented in the \texttt{smacofSphere} function of the \texttt{smacof} R package \citep{MDS-Majorization}. Both SMDS algorithms were initiated by a projection to $\mathbb{S}^d$ of the classical MDS solution. Calculation times for the optimization of \eqref{eq:smds3} also include the initialization of the radius according to \eqref{eq:wasser}. Optimization in \eqref{eq:smds3} was conducted at a relative tolerance level of $5\times 10^{-2}$ and MDS-Q at $\varepsilon = 5\times 10^{-2}$, which was chosen due to comparable computational efficiency and notable improvements in stress over MDS-Q and can partly be attributed to MDS-Q's non-geodesic calculations of distances. For the dual MDS-Q algorithm, the default penalty (10) was used. Each of the three numerical examples in Section \ref{sec:simul} was generated 50 times and their results were averaged. In the setting of Subsection \ref{subsec:simul_2D}, our optimization of \eqref{eq:smds3} took 6.45s and yielded a stress of 0.068, primal MDS-Q needed 14.29s and yielded a stress of 0.479, while dual MDS-Q reported 2.16s and 1.076, respectively. The respective metrics for the setting of Subsection \ref{subsec:simul_3D} are: 7.28s and 0.064; 75.02s (with a range of 43--157s) and 1.23; 2.6 and 1.753. For the setting in Subsection \ref{subsec:simul_2D_wrapped} we obtained: 30.93s and 0.159; 14.65s and 0.645; and 2.19s and 0.691. Due to the above considerations, and since we specifically target geodesic distances, we have decided to use approach \eqref{eq:smds3} in the sequel.

%---------------------------%
\subsection{Considerations on the radius}
\label{subsec:radius}
%---------------------------%

The choice of $r$ in \eqref{eq:smds} and \eqref{eq:smds2} affects the scaling of the spherical distances. This has an important effect, since distances on $\mathbb{S}^d_r$ are bounded by $r \pi$ while distances on $\mathbb{T}^d$ are bounded by $\sqrt{d} \pi$. Besides, $\mathbb{S}^d_r$ becomes a very different space when $r\to0$ or $r\to\infty$, collapsing in the first case to $\{\mathbf{0}\}$ and becoming flat in the second limit case (the curvature of $\mathbb{S}_r^d$ is $1/r^2$).\\

A possibility is to incorporate $r$ into the optimization problem, modifying the radius of the arrival space in a data-driven form, as done by \cite{MDS-Optimization}. Precisely, we can turn \eqref{eq:smds2} into
\begin{align}
	(\widehat{r},\widehat{\bm{z}}_1,\ldots,\widehat{\bm{z}}_n) = \underset{(r,\bm{z}_1,\ldots,\bm{z}_n)\in \mathbb{R}_+\times(\mathbb{R}^{d+1})^n}{\arg\min} \frac{1}{n(n-1)}
	\sum_{i \neq j } \left(\delta_{\mathbb{T}^d}(\bm{x}_i,\bm{x}_j) - r\arccos\left( \frac{\bm{z}_i'\bm{z}_j}{ \|\bm{z}_i\|\|\bm{z}_j\|}\right)\right)^2.
	\label{eq:smds3}
\end{align}
Including $r$ within the optimization in \eqref{eq:smds2} is immediate, which is an additional benefit of \eqref{eq:smds2}. A simple initial value for $r$ is $\sqrt{d}$; it equates the maximum distances on $\mathbb{T}^d$ and $\mathbb{S}_r^d$.\\

% Initial values
A more sophisticated approach to select $r$ that does not require expanding \eqref{eq:smds2} to \eqref{eq:smds3} is described next. Denote by $\mu$ and $\nu_r$ the two uniform measures on $\mathbb{T}^d$ and $\mathbb{S}_r^d$, respectively. Let $\bm{u}_1,\ldots,\bm{u}_n\sim \mu$ and $\bm{v}_1,\ldots,\bm{v}_n\sim \nu_r$ be iid samples. Denote by $F_{N,\mathbb{T}^d}$ to the empirical cumulative distribution function (ecdf) of the pairwise distances $\{\delta_{\mathbb{T}^d}(\bm{u}_i,\bm{u}_j)\}_{1\leq i<j\leq n}$ and by $G_{N,\mathbb{S}_r^d}$ the analogous ecdf for $\{\delta_{\mathbb{S}_r^d}(\bm{v}_i,\bm{v}_j)\}_{1\leq i<j\leq n}$, with $N:=n(n-1)/2$. Define
\begin{align}
	r^* := \arg \min_{r \geq 0} \mathbb{E}_{\mu\times\nu_r}\left[W_1(F_{N,\mathbb{T}^d}, G_{N,\mathbb{S}_r^d})\right], \quad W_1(F_{N,\mathbb{T}^d}, G_{N,\mathbb{S}_r^d})=\int_0^1 \big|F^{(-1)}_{N,\mathbb{T}^d} (s) - G^{(-1)}_{N,\mathbb{S}_r^d} (s) \big| \,\mathrm{d}s,\label{eq:wasser}
\end{align}
which corresponds to the expected $1$-Wasserstein distance between the two pairwise distance ecdfs under the product measure $\mu\times\nu_r$. Above, $F^{(-1)}$ stands for the of generalized inverse of $F$. As defined, $r^*$ guarantees that, on average, the most spread sample of size $n$ on $\mathbb{T}^d$ can be allocated on $\mathbb{S}^d_r$ while optimally maintaining its interpoint-distance distribution. Uniformly-distributed samples on $\mathbb{T}^d$ can be regarded as the worst-case scenario on which to perform SMDS effectively; $r^*$ is designed to aid precisely in this situation. In practice, the expectation in \eqref{eq:wasser} can be estimated empirically by Monte Carlo and the Wasserstein distance is evaluated by its equivalent form $W_1(F_{N,\mathbb{T}^d}, G_{N,\mathbb{S}_r^d})=\frac{1}{N} \sum_{i=1}^N \big|d_{(i),\mathbb{T}^d} -d_{(i),\mathbb{S}_r^d} \big|$ for $d_{(k),X}$ being the $k$-th order statistic of $\{\delta_{X}(\bm{u}_i,\bm{u}_j)\}_{1\leq i<j\leq n}$. In the experiments described in Subsection \ref{subsec:implem}, as well as in the sequel, we computed $r^*$ as described in \eqref{eq:wasser} estimating the expectation with $M=100$ Monte Carlo replicates of $n=100$ uniformly generated points each. Then, we have used this as an initial value for $\widehat{r}$ in solving \eqref{eq:smds3}.

%---------------------------%
\subsection{PNS fit}
\label{subsec:pns}
%---------------------------%

% Now we get to PNS
By solving \eqref{eq:smds2} we have found a configuration $\widehat{\bm{y}}_1,\ldots,\widehat{\bm{y}}_n \in \mathbb{S}^d_r$. For the easiness of presentation, we assume this configuration has been projected to $\mathbb{S}^d$ before applying PNS (which is invariant to scaling). We can now use PNS on $\widehat{\bm{y}}_1,\ldots,\widehat{\bm{y}}_n$ to obtain a nested sequence of subspaces
\begin{align} \label{eq:PNS}
	\mathbb{S}^d\supset S^{d-1} \supset S^{d-2} \supset \ldots \supset S^1 \supset \{\widehat{\boldsymbol\mu}\}, 
\end{align}
where $S^{d-1} \cong \mathbb{S}^{d-1}, \ldots, S^1 \cong \mathbb{S}^1$, and $\widehat{\boldsymbol\mu}$ is the \textit{backwards mean}, which is the Fréchet mean of the projections of the data onto $S^1$. A possible parametrization for the $(d-1)$-dimensional subsphere $S^{d-1}$ that sheds light into its small-subsphere structure is attained with the tangent-normal decomposition: $S^{d-1}=\{\bm{x}\in\mathbb{S}^d:\bm{x}=t_1\boldsymbol\theta_1+(1-t_1^2)^{1/2}\bm{B}_{\boldsymbol\theta_1}\boldsymbol\xi_1,\,\boldsymbol\xi_1\in\mathbb{S}^{d-1}\}$, where $(t_1,\boldsymbol\theta_1)\in[-1,1]\times\mathbb{S}^{d}$ and $\bm{B}_{\boldsymbol\theta_1}$ is an arbitrary semi-orthonormal $(d+1)\times d$ matrix such that $\bm{B}_{\boldsymbol\theta_1}\bm{B}_{\boldsymbol\theta_1}'=\bm{I}_{d+1}-\boldsymbol\theta_1\boldsymbol\theta_1'$ and $\bm{B}_{\boldsymbol\theta_1}'\bm{B}_{\boldsymbol\theta_1}=\bm{I}_{d}$. Further subspheres can be parametrized iteratively in a nested fashion, yet explicit forms are more convoluted.\\

% Space of Principal Scores
PNS also yields the matrix of scores, $\{ \bm{\xi}_i\}_{i = 1}^n$, where $\bm{\xi}_i$ is a $d$-vector (for angular parametrization) that denotes the coordinates of $\widehat{\bm{y}}_i$ with regard to the spherical components. This matrix gives rise to the space of principal scores $E = [-\pi,\pi) \times [-\pi/2,\pi/2]^{d-1}$, a map $h: \mathbb{S}^d \to E$ of PNS such that $h(\widehat{\bm{y}}_i) = \bm{\xi}_i$, and its inverse, $\tilde{h}$. Further, we denote by $\{\bm{\xi}_i^{(k)} \}$ the projection of the $i$-th datum to $S^{k-1}$, so that $\bm{\xi}_i^{(k)}$ agrees with $\bm{\xi}_i$ in the first $k$-coordinates and is $0$ in the rest. Essentially, given a $\boldsymbol \xi \in E$, where $ \xi_j$ denotes the deviation from the $j$-th nested sphere (of dimension $d-j$),
\begin{align*}
	\tilde{h} (\boldsymbol \xi)  = (\tilde{g}_{1} \circ \cdots \circ\tilde{g}_{d-1} )(\boldsymbol \xi), \quad \tilde{g}_{k}(\boldsymbol \xi^{(k-1)}, \xi_k) := \bm{R}(\bm{v}_k)'\begin{pmatrix}\sin(r_k + \xi_k)\boldsymbol \xi^{(k-1)}\\\cos(r_k + \xi_k)\end{pmatrix}, \quad k = 1,\ldots, d,
\end{align*}
where $\bm{R}(\bm{v}_k)$ is the matrix that rotates $\bm{v}_k$, the center of the $k$-th subsphere, to the north pole, and $r_k$ its radius.\\

% Implementation and Further Reading
For more information on the maps $h$ and $\tilde{h}$ the reader is referred to Section 4 of the Supplementary Material of \cite{PNS} and to the main paper for the construction of the nested spheres. PNS is implemented in the \texttt{pns} function of the \texttt{shapes} R package \citep{shapes}, which we have used.

%---------------------------%
\subsection{Prediction}
\label{subsec:pred}
%---------------------------%

Step \ref{step1} of ST-PCA grants an association between the point configurations $(\bm{x}_1,\ldots,\bm{x}_n) \in (\mathbb{T}^d)^n$ and $(\widehat{\bm{y}}_1,\ldots,\widehat{\bm{y}}_n)\in(\mathbb{S}^d_r)^n$. Suppose that we are given an arbitrary point $\bm{y} \in \mathbb{S}_r^d$ and are asked to \textit{predict} a configuration $\bm{x}\in \mathbb{T}^d$ that best preserves the interpoint distances of $\bm{x}$ with respect to the original configuration $\bm{x}_1,\ldots,\bm{x}_n$. This problem is referred to as the \emph{prediction problem} for MDS, see, e.g., \cite{Biplots}. This ``prediction'' can be achieved by defining the set-valued map $\widetilde{T}:\mathbb{S}^d_r\to\mathbb{T}^d$ given by
\begin{align}
	\widetilde{T}(\bm{y})
	:=\arg\min_{\bm{x}\in\mathbb{T}^d}\frac{1}{n}\sum_{i=1}^n\left(\delta_{\mathbb{T}^d}(\bm{x},\bm{x}_i)-\delta_{\mathbb{S}_r^d}(\bm{y},\widehat{\bm{y}}_i)\right)^2,
	\label{eq:invmds2}
\end{align}
which is a \emph{coherent} prediction problem \citep{Biplots} since the same objective function (stress) as in the original SMDS problem is used. In practice, an appropriate initialization for the numerical optimization to be done in \eqref{eq:invmds2} is the sample point $\bm{x}_i$ whose $\widehat{\bm{y}}_i$ is closest to $\bm{y}$. For the sake of completeness, we mention that the \emph{interpolation problem} for MDS refers to the reverse problem to \eqref{eq:invmds2}, which induces the definition of the set-valued map $\widehat{T}:\mathbb{T}^d\to\mathbb{S}^d_r$ given by
\begin{align}
	\widehat{T}(\bm{x}):=\arg\min_{\bm{y}\in\mathbb{S}_r^d}\frac{1}{n}\sum_{i=1}^n\left(\delta_{\mathbb{T}^d}(\bm{x},\bm{x}_i)-\delta_{\mathbb{S}_r^d}(\bm{y},\widehat{\bm{y}}_i)\right)^2.\label{eq:interp}
\end{align}
Both \eqref{eq:invmds2} and \eqref{eq:interp} give a means of extending the SMDS matching between $(\bm{x}_1,\ldots,\bm{x}_n)$ and $(\widehat{\bm{y}}_1,\ldots,\widehat{\bm{y}}_n)$ to arbitrary points on $\mathbb{T}^d$ and $\mathbb{S}_r^d$, respectively.\\

We now explain how prediction is used to find a tractable first principal component on $\mathbb{T}^d$. Recall from Section \ref{subsec:pns} that PNS yields an explicit map, $\Tilde{h}:E \rightarrow \mathbb{S}^d$. Using $\Tilde{h}$ we can obtain the Euclidean parametrization of $S^1$ in $\mathbb{S}^d$. The descriptor is obtained in two steps:
\begin{enumerate}
	\item Take a grid $\{\xi_j\}_{j=1}^m\subset[-\pi,\pi)$. Then, $\{\boldsymbol \xi_j\}_{j=1}^m \subset E$ is a discrete representation of $S^1$, where $\boldsymbol \xi_j = \xi_j \times \{ 0 \} \times \stackrel{d-1}{\cdots} \times\{ 0 \}$, and obtain $\{\boldsymbol \psi_j\}_{j=1}^m$ for $\boldsymbol \psi_j:=\Tilde{h}(\boldsymbol \xi_j)$.
	\item Compute $\widetilde{T}(\boldsymbol \psi_j)$ for each $j=1,\ldots,m$ by numerically optimizing \eqref{eq:invmds2}. To aid continuity in the resulting sequence of solutions of the optimization problems and avoid spurious minima, use the former prediction $\widetilde{T}(\boldsymbol \psi_j)$ as starting value for obtaining the next one $\widetilde{T}(\boldsymbol \psi_{j+1})$. 
\end{enumerate} 
The above procedure generates a discretization of the set
\begin{align*}
	\widetilde{S}^1: = \widetilde{T}(S^1) = \{ \widetilde{T} (\boldsymbol\psi) \in\mathbb{T}^d:\boldsymbol \psi\in S^1 \} .
\end{align*}

We close this subsection with a remark. The strength and flexibility of ST-PCA relies on the fact that it searches for a transformation that is, in some sense, optimal. Since this is a solution of optimization problems, there is no guaranteed regularity for ST-PCA's principal components on the torus. This can become apparent when predicting $S^1$, a continuous curve on $\mathbb{S}^d$ whose predicting curve may yield discontinuities or non-closed curves in data-sparse regions of $\mathbb{T}^d$. In practice, the continuity of $\widetilde{S}^1$ can be facilitated by sensible initializations, as above described, followed by a fine-grain reconstruction of $\widetilde{S}^1$ by linear interpolations of $\{\widetilde{T}(\boldsymbol \psi_{j})\}_{j=1}^m$ that are adapted to $\mathbb{T}^d$. Nevertheless, for purposes of clustering and data visualization, Steps \ref{step1} and \ref{step2} of our algorithm suffice. Hence, even in the case that the principal component constructed by Step \ref{step3} of ST-PCA is discontinuous, this algorithm can yield meaningful dimensionality reduction.

%---------------------------%
\subsection{Explained variance}
\label{subestction:var_explained}
%---------------------------%

Consider the projection of $\widehat{\bm{y}}_i$ to $S^k$, denoted by $\widehat{\bm{y}}_i^{(k)}$, $k = 0,\ldots,d$, where $\widehat{\bm{y}}_i^{(0)} = \widehat{\boldsymbol{\mu}}$, $i = 1,\ldots, n$, and $\widehat{\boldsymbol{\mu}}$ is the Fréchet mean of the projections of the data on $S^1\cong\mathbb{S}^1$. Then, the variance decomposition provided by PNS for such sample is
\begin{align}
	\mathrm{var}_{\mathbb{S}^d}(\widehat{\bm{y}}_1,\ldots,\widehat{\bm{y}}_n) :=\sum_{k=1}^d \mathrm{var}_{\mathbb{S}^d}^{(k)}(\widehat{\bm{y}}_1,\ldots,\widehat{\bm{y}}_n):=  \sum_{k=1}^{d}\sum_{i=1}^n\left(\delta_{\mathbb{S}^d} \big(\widehat{\bm{y}}_i^{(k-1)},\widehat{\bm{y}}_i^{(k)}\big)\right)^2 \label{eq:totalvar}
\end{align}
and the proportion of variance explained by the $k$-th component is $\mathrm{var}_{\mathbb{S}^d}^{(k)}(\widehat{\bm{y}}_1,\ldots,\widehat{\bm{y}}_n)/\allowbreak\mathrm{var}_{\mathbb{S}^d}(\widehat{\bm{y}}_1,\ldots,\widehat{\bm{y}}_n)$.\\

Denote by $\widehat{\bm{x}}_i^{(k)} = \widetilde{T}(\widehat{\bm{y}}_i^{(k)})$, $k = 0,\ldots,d$, to the predicted projection $\widehat{\bm{y}}_i^{(k)}$, which can be equivalently regarded as the projection of $\bm{x}_i$ to $\widetilde{S}^k$. Then, based on \eqref{eq:totalvar}, we define the total variance of $\bm{x}_1,\ldots,\bm{x}_n$ on $\mathbb{T}^d$ as
\begin{align*}
	\mathrm{var}_{\mathbb{T}^d}(\bm{x}_1,\ldots,\bm{x}_n) := \sum_{k=1}^{d} \mathrm{var}^{(k)}_{\mathbb{T}^d}(\bm{x}_1,\ldots,\bm{x}_n) := \sum_{k=1}^{d}\sum_{i=1}^n  \left(\delta_{\mathbb{T}^d} \big(\widehat{\bm{x}}_i^{(k-1)},\widehat{\bm{x}}_i^{(k)}\big)\right)^2.
\end{align*}
The proportion of variance explained by the $k$-th component is then defined as $\mathrm{var}_{\mathbb{T}^d}^{(k)}(\bm{x}_1,\ldots,\bm{x}_n)/\allowbreak\mathrm{var}_{\mathbb{T}^d}(\bm{x}_1,\ldots,\bm{x}_n)$. In particular, this ratio allows quantifying the proportion of data variance that is captured by $\widetilde{S}^1$, a popular metric of success in dimensionality reduction. As it is typical for PCA analogs in non-Euclidean spaces \citep[see, e.g., the discussion in Section 2.3 in][]{Eltzner2018}, there is no guarantee that $k\mapsto \mathrm{var}_{\mathbb{T}^d}^{(k)}(\bm{x}_1,\ldots,\bm{x}_n)/\allowbreak\mathrm{var}_{\mathbb{T}^d}(\bm{x}_1,\ldots,\bm{x}_n)$ is a non-decreasing function.

%---------------------------%
\section{Numerical experiments}
\label{sec:simul}
%---------------------------%

%---------------------------%
\subsection{Clusters toy data on \texorpdfstring{$\mathbb{T}^2$}{T\textasciicircum2}}
\label{subsec:simul_2D}
%---------------------------%

We demonstrate next the ability of ST-PCA to separate clustered data on $\mathbb{T}^2$ without distorting the cluster structure. To that aim, a toy dataset consisting of three isotropic clusters with means $\boldsymbol\mu_1 = (-1,-2)'$, $\boldsymbol\mu_2 = (3,0.5)'$, and $\boldsymbol\mu_3 = (-0.8,2.5)'$, each with $n=100$, has been considered. The dataset is displayed in Figure \ref{fig:T2_TORUS}. We performed Step \ref{step1} to obtain a configuration of the data on $\mathbb{S}^2$, displayed in Figure \ref{fig:T2_spherical_view}. For this specific dataset, we have obtained $r^* = 1.47$ as an initial value from \eqref{eq:wasser}, which is then modified to $\widehat{r} = 1.35$ through \eqref{eq:smds3}. In Step \ref{step2}, we obtained the PNS scores plot for this spherical configuration, displayed in Figure \ref{fig:T2_SPHERE}. The black curve in Figure \ref{fig:T2_spherical_view} shows $S^1$, the best fitting $\mathbb{S}^1$ to the spherical data, which corresponds to the horizontal axis of the scores plot in Figure \ref{fig:T2_SPHERE}.

\begin{figure}[!h]
	\vspace*{-0.65cm}
	\centering
	\subfloat[]{\includegraphics[width=0.33\textwidth]{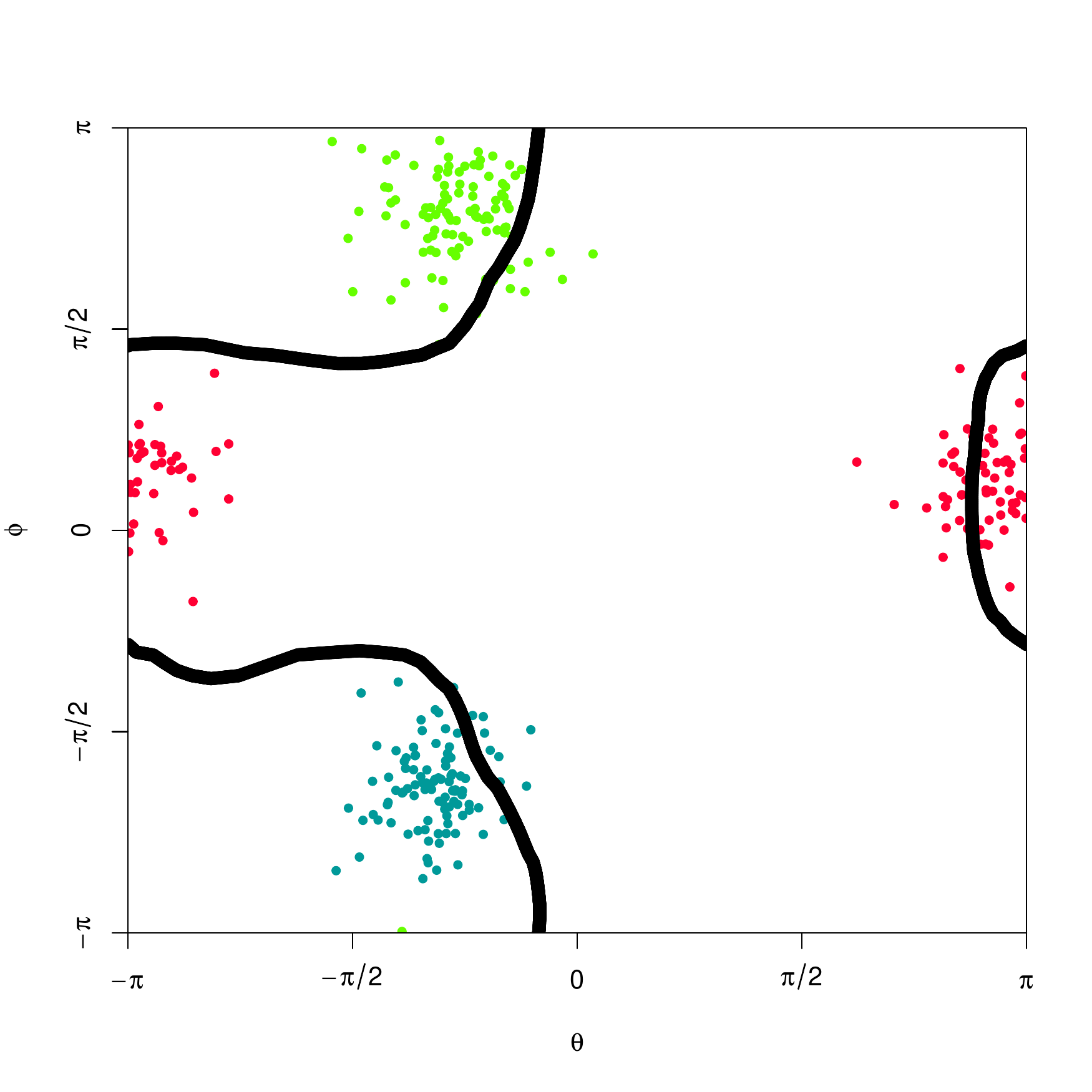} \label{fig:T2_TORUS}}
	\hfill
	\subfloat[]{\includegraphics[width=0.33\textwidth]{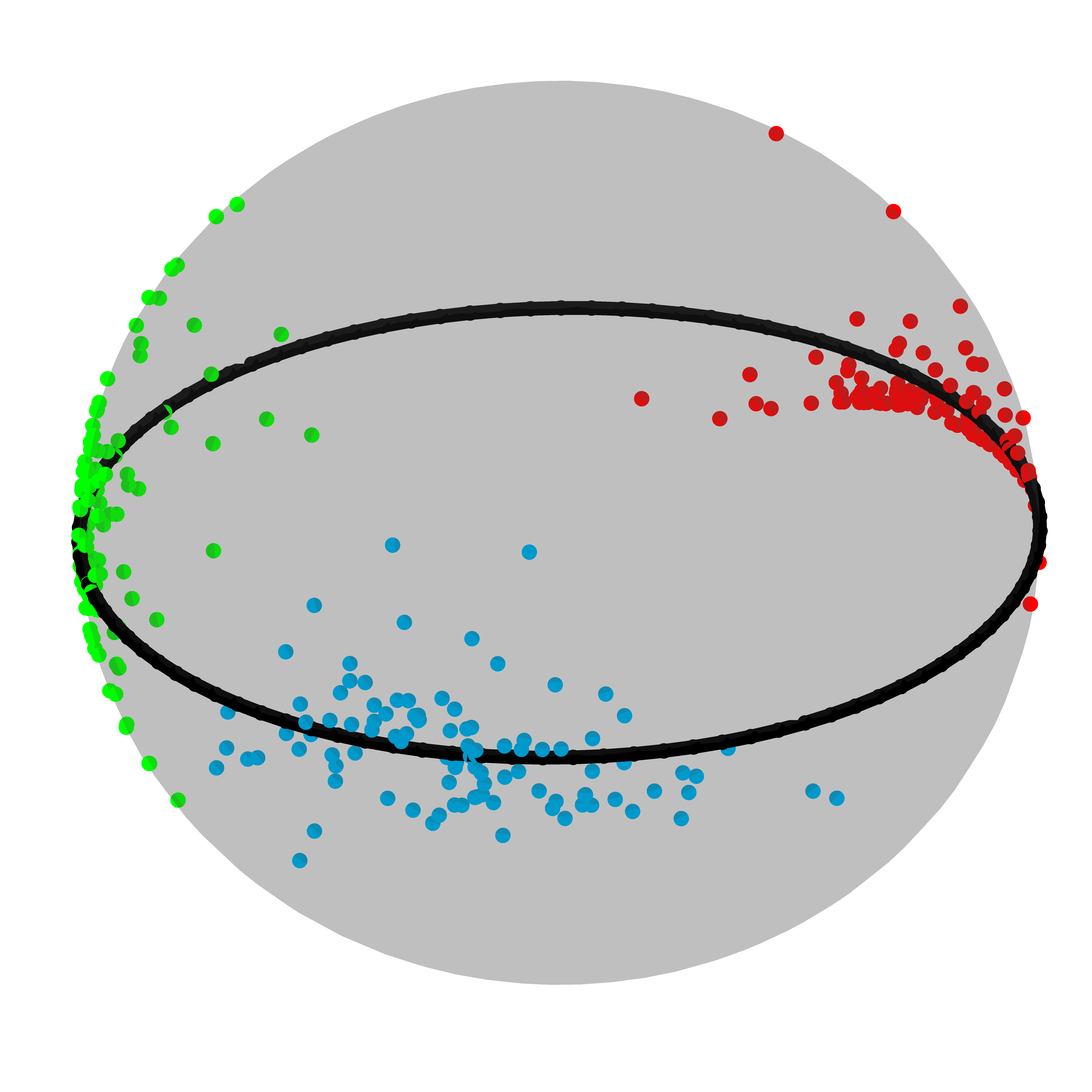}\label{fig:T2_spherical_view}}
	\subfloat[]{\includegraphics[width=0.33\textwidth]{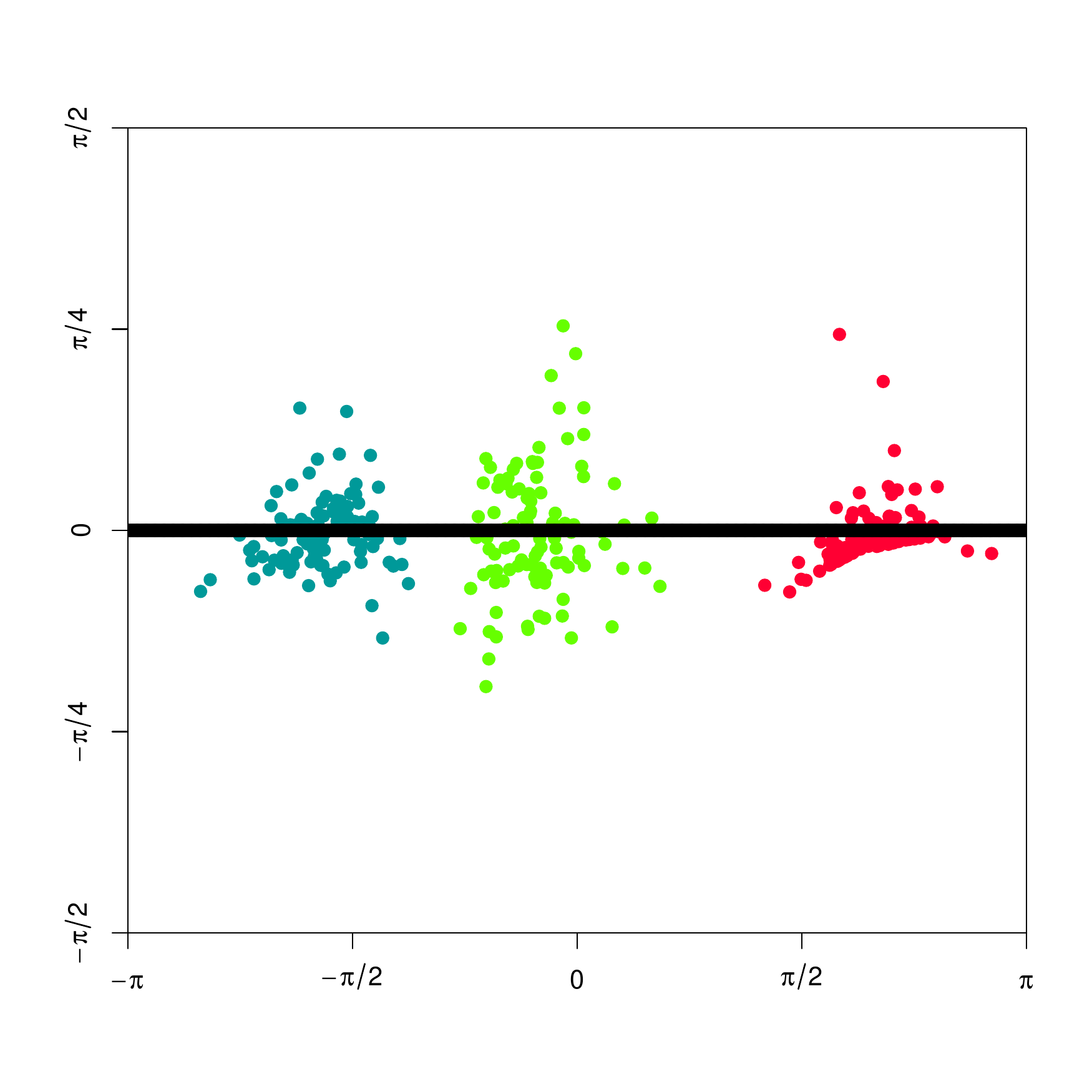}\label{fig:T2_SPHERE}}
	\hfill
	\caption{\small Panel (a) shows the three clusters of simulated data on $\mathbb{T}^2$. Panel (b) presents the sample SMDS-mapped to $\mathbb{S}^2$. Panel (c) displays the PNS scores. The principal mode of variation on the torus, sphere, and scores is displayed in all panels as a black curve. The first ST-PCA mode of variation clearly captures the three clusters.}
	\label{fig:Simulation_T2}
\end{figure}

The resulting $\widetilde{S}^1$, computed in Step \ref{step3} of ST-PCA, is shown in the black curve in Figure \ref{fig:T2_TORUS}. The figure reveals that the first toroidal component efficiently adapts to the clustered structure of the data (percentage of variance explained: $95\%$). $\widetilde{S}^1$ was obtained by predicting an equispaced grid $\{\boldsymbol\psi_j\}_{j=1}^{100}\subset S^1$ and then linearly interpolating the resulting predictions on $\mathbb{T}^2$ to yield a smooth~curve.

%---------------------------%
\subsection{Clusters toy data on \texorpdfstring{$\mathbb{T}^3$}{T\textasciicircum3}}
\label{subsec:simul_3D}
%---------------------------%

We now simulate data from an isotropic mixture of three wrapped normal distributions in $\mathbb{T}^3$. We simulated $n=300$ points equally distributed in the three components with centers $\boldsymbol\mu_1 = (-1,-2,0)'$, $\boldsymbol\mu_2 = (1,1,-1)'$, and $\boldsymbol\mu_3 = (0,-1,3)'$. Figure \ref{fig:T3_TORUS} displays the three clusters on $\mathbb{T}^3$, where each side of the cube is glued to the opposite side. After running SMDS in Step \ref{step1}, we have obtained a configuration on points $\widehat{\bm{y}}_1,\ldots,\widehat{\bm{y}}_{n} \in \mathbb{S}^3$. For visualization purposes, we show in Figure \ref{fig:T3_spherical_view} the spherical view of the configuration's projections to $S^2\cong\mathbb{S}^2$. For this dataset we have obtained $r^*=1.79$, which then yielded $\widehat{r} = 1.63$. Figure \ref{fig:T3_SPHERE} shows the scores plot after applying PNS to $\widehat{\bm{y}}_1,\ldots,\widehat{\bm{y}}_{n} \in \mathbb{S}^3$ (Step \ref{step2}). Finally, we inverted $S^1$, the black curve in Figure \ref{fig:T3_spherical_view}, to produce $\widetilde{S}^1$ as displayed in the black curve of Figure \ref{fig:Simulation_T3}. We did so with an equispaced grid $\{\boldsymbol\xi_j\}_{j=1}^{100}\subset S^1$ complemented by a periodic linear interpolation on $\mathbb{T}^3$. The entire cluster structure is captured by the first mode of variation, which clearly exploits the periodicity of $\mathbb{T}^3$ for efficiently explaining the data variability. The percentage of variance explained by the first and second ST-PCA components is $94\%$ and $1\%$, respectively.

\begin{figure}[!h]
	\vspace*{-0.5cm}
	\centering
	\subfloat[]{\includegraphics[width=0.33\textwidth]{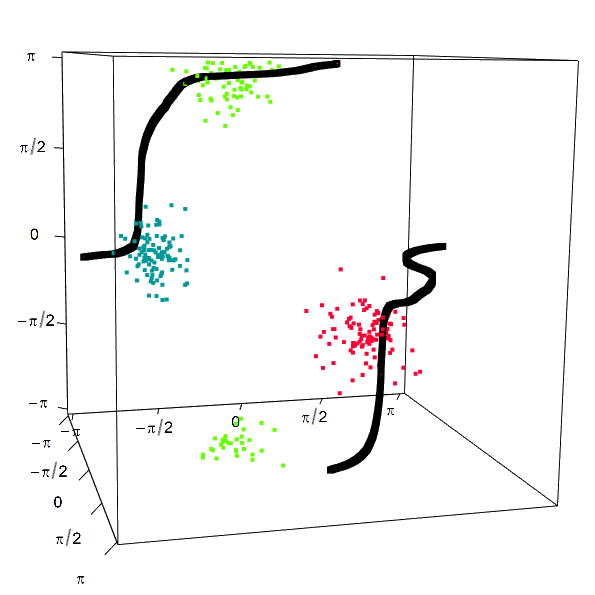} \label{fig:T3_TORUS}}
	\hfill
	\subfloat[]{\includegraphics[width=0.33\textwidth]{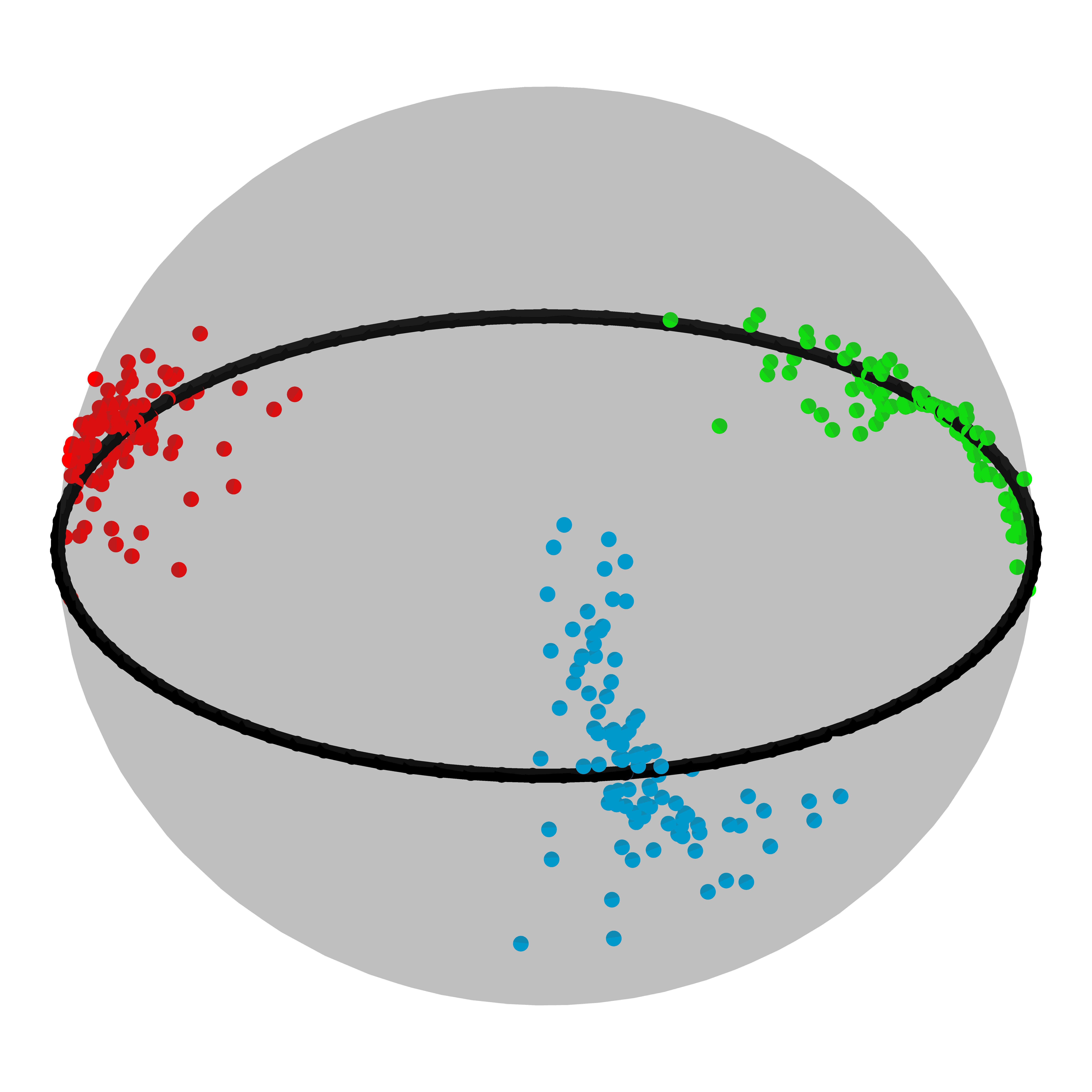}\label{fig:T3_spherical_view}}
	\subfloat[]{\includegraphics[width=0.33\textwidth]{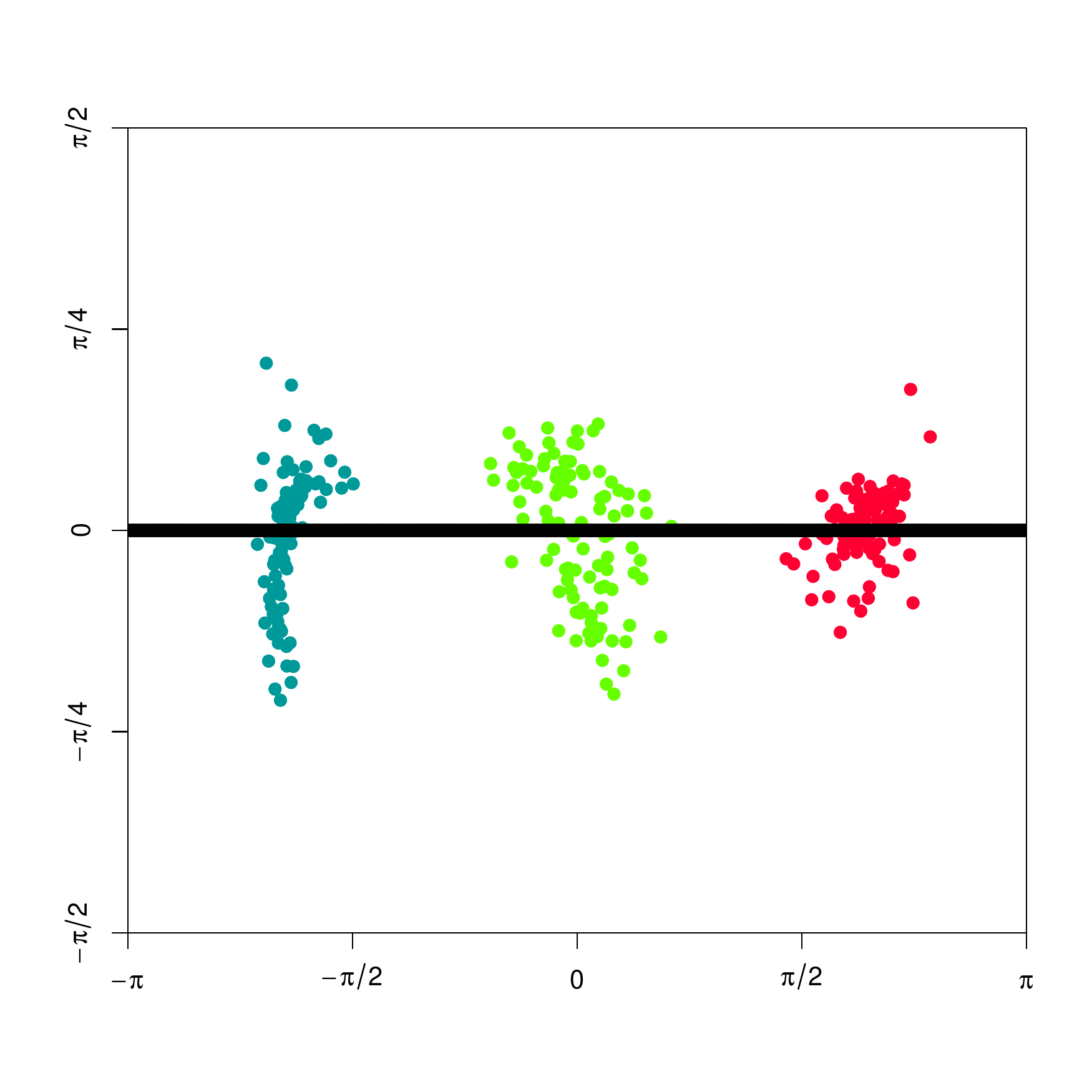}\label{fig:T3_SPHERE}}
	\hfill
	\caption{\small Panel (a) shows the three clusters of simulated data on $\mathbb{T}^3$. Panel (b) presents the projection to $S^2\cong\mathbb{S}^2$ of the sample SMDS-mapped to $\mathbb{S}^3$. Panel (c) displays the first PNS scores. The principal mode of variation on the torus, sphere, and scores is displayed in all panels as a black curve. The first ST-PCA principal component successfully separates the three clusters.}
	\label{fig:Simulation_T3}
\end{figure}

%---------------------------%
\subsection{Anisotropic wrapped normal on \texorpdfstring{$\mathbb{T}^2$}{T\textasciicircum2}}
\label{subsec:simul_2D_wrapped}
%---------------------------%

In the final numerical experiment we simulated $n=500$ observations from a wrapped normal distribution on $\mathbb{T}^2$ with mean $\boldsymbol\mu = (-1,0)'$ and covariance matrix $\boldsymbol \Sigma = (2, 2; 2, 3)$. Figure \ref{fig:Simulation_T2_wrapped} shows the analogous analysis to that carried out in Figure \ref{fig:Simulation_T2}, this time using a rainbow palette for easier identification of the first ST-PCA principal component and with $r^*=1.48$ and $\widehat{r} = 1.39$. The mode of variation successfully utilizes the periodicity of the torus to optimally capture the data variability, attaining a percentage of variance explained of $89\%$.

\begin{figure}[!h]
	\vspace*{-0.5cm}
	\centering
	\subfloat[]{\includegraphics[width=0.33\textwidth]{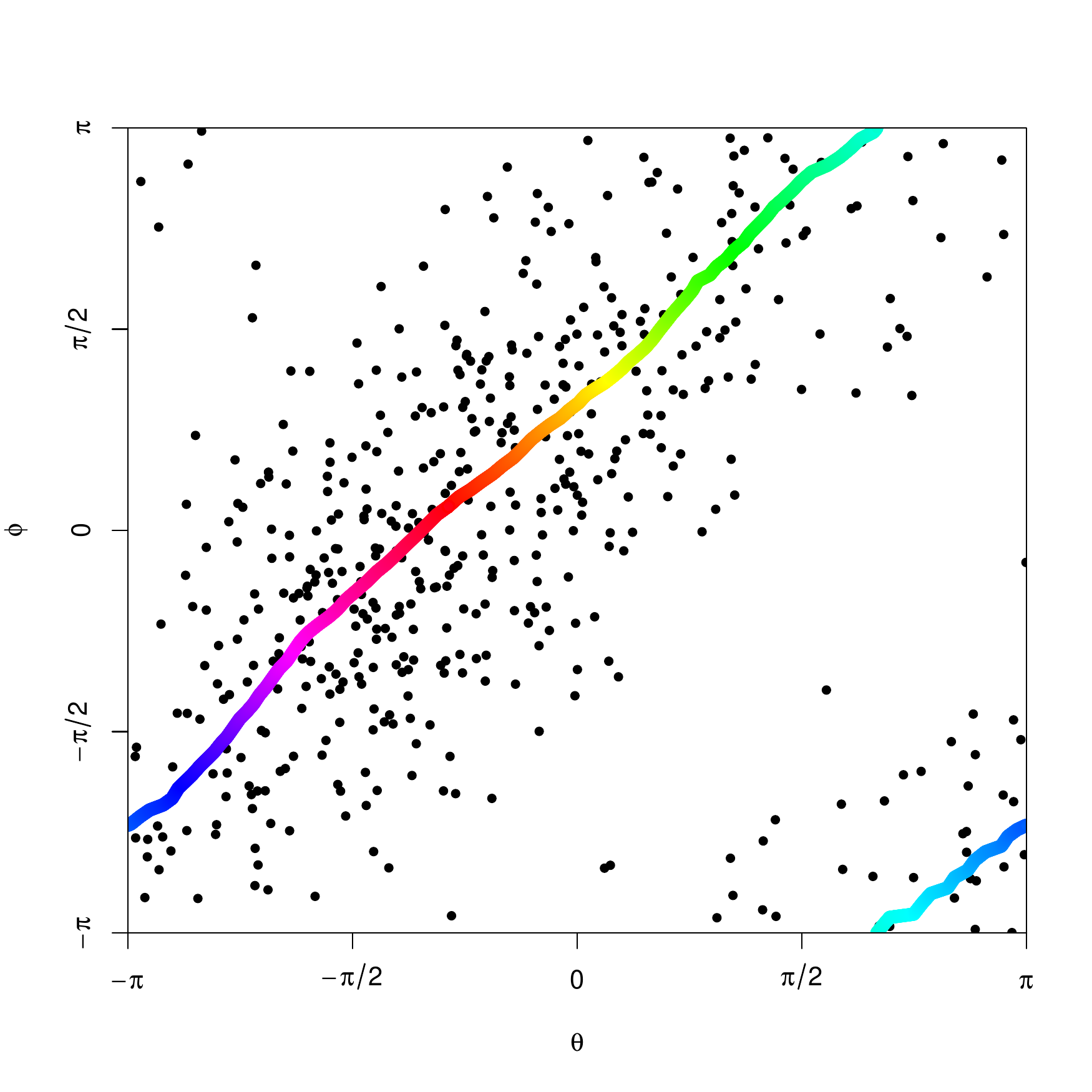} \label{fig:T2_wrapped_TORUS}}
	\hfill
	\subfloat[]{\includegraphics[width=0.33\textwidth]{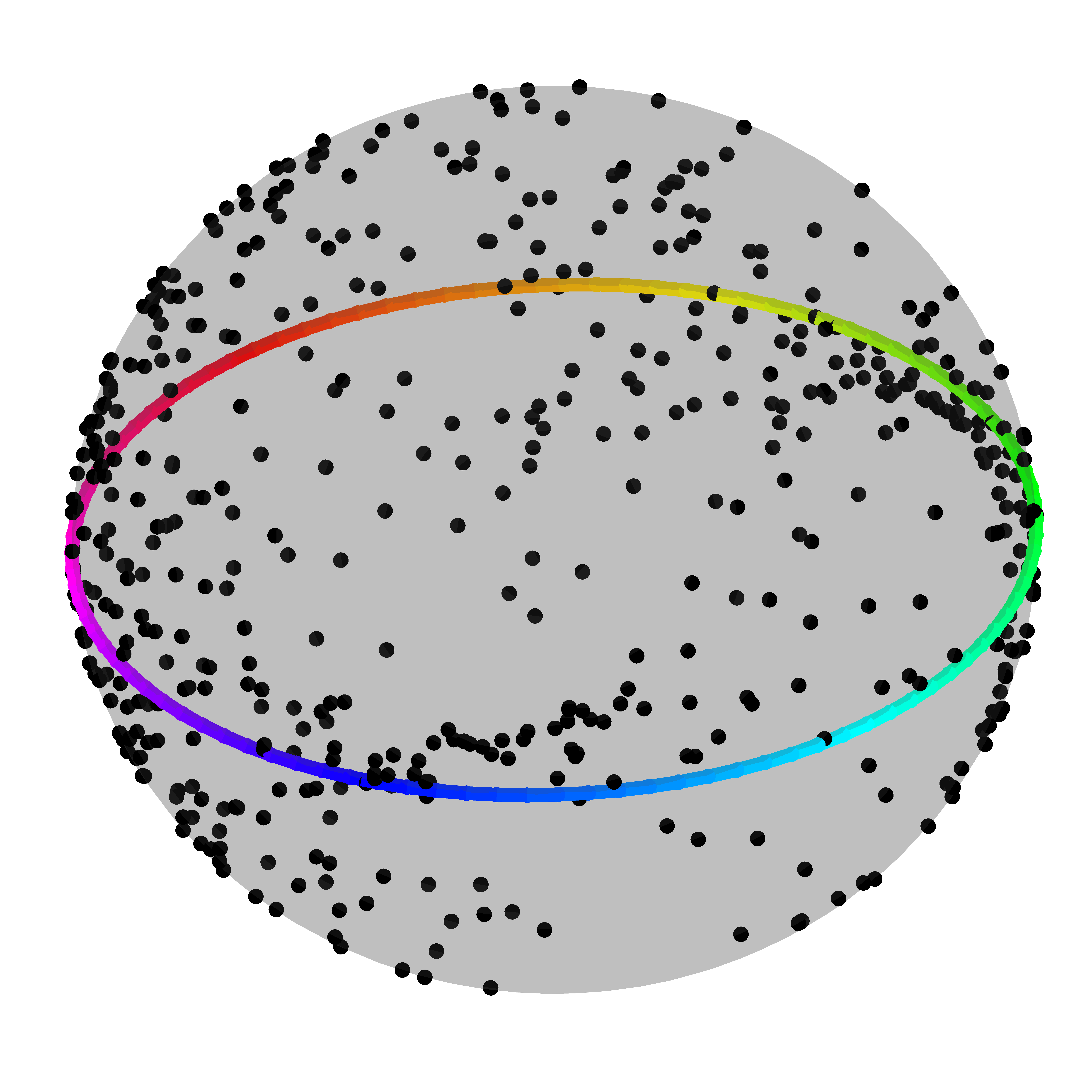}\label{fig:T3_wrapped_spherical_view}}
	\subfloat[]{\includegraphics[width=0.33\textwidth]{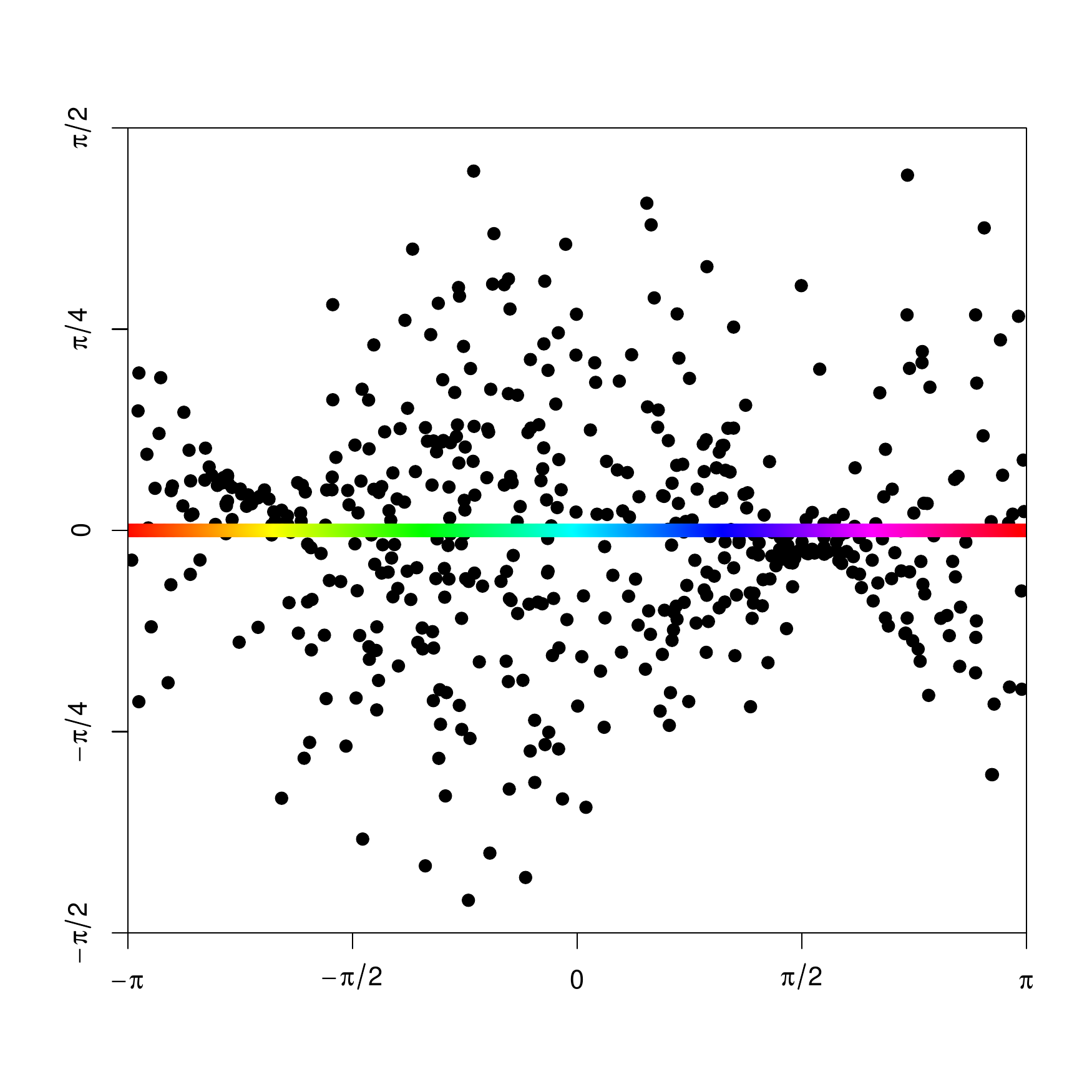}\label{fig:T3_wrapped_SPHERE}}
	\hfill
	\caption{\small Panel (a) shows the sample from a wrapped Gaussian distribution on $\mathbb{T}^2$ ; panel (b): spherical view of the configuration of points obtained by MDS. Panel (c) displays the first PNS scores. The principal mode of variation from the sphere is displayed in all panels in a rainbow palette. The first ST-PCA mode of variation spans along the presumed center of the banded data-dense region.}
	\label{fig:Simulation_T2_wrapped}
\end{figure}

%---------------------------%
\section{Real data applications}
\label{sec:real}
%---------------------------%

%---------------------------%
\subsection{Sunspots}
\label{subsec:Sunspots}
%---------------------------%

% Sunspots
Sunspots are cooler regions of the Sun's photosphere that are related with solar magnetic field concentrations. They have been used as a measure of the Sun's solar activity and are speculated to have an impact on Earth's long-term climate \citep[see, e.g.,][]{Haigh}. The solar magnetic field generating the sunspots varies according to a nearly periodic, 11-year period that is referred to as a solar cycle. Sunspots observations have been well documented in the Debrecen Photoheliographic Data and the Greenwhich Photoheliographic Results. From the latter source, \cite{rotasym} give a curated dataset with the centers of groups of sunspots. The locations of the centers refer to the  first-ever observations of such sunspots (henceforth referred to as ``births''). We focus on the last cycle that has been fully observed with curated data, the \textit{23rd solar cycle}. It includes a total of 5373 records ranging from August of 1996 to November of 2001. Even though the main generator of sunspots, the twisting of the solar magnetic field, has a rotationally symmetric nature, the existence of \textit{preferred longitudes} has been speculated for the recurrent appearance of sunspots \citep{Babcock1961,Bogart1982,Pelt2010}.\\

% Case study
We explore the dependence structure of the longitudes $\theta\in [-\pi,\pi)$ of sunspots births. Since the Earth rotates around the Sun, sunspots visibility is limited by which side of the Sun is facing the Earth. Therefore, if there existed significantly preferred longitudes, a significant deviation from a linear dependence structure in the series of sunspots longitudes should be expected. To investigate this serial structure, we jointly consider the series of longitudes and its lagged versions of order one and two,
\begin{align*}
	\boldsymbol\Theta_i := (\theta_i,\theta_{i+1},\theta_{i+2})' \in \mathbb{T}^3, \quad i = 1, \ldots, 5371,
\end{align*}
the main driver behind that selection being showing graphically the output of ST-PCA on $\mathbb{T}^3$. The principal mode of variation for the longitudes of the births of three consecutive groups of sunspots is shown in Figure \ref{fig:STPCA-TPCA_Variation}. The figure reveals that the data are concentrated around the (periodic) diagonal line and that ST-PCA (with $\widehat{r}=1.81$) correctly identifies this line as the first mode of variation, $\widetilde{S}^1$. Recall that the points on the edges of the cube belong to the cloud of points around the diagonal, thus rendering it as the most sensible choice for a principal mode of variation. $\widetilde{S}^1$ is remarkably linear: a suitably-adapted linear fit to the grid of 100 points defining $\widetilde{S}^1$ yields $R^2=1-3\times 10^{-5}$. Therefore, the apparent linearity of the innovations in the series of sunspots births longitudes points towards the non-existence of (at least) major preferred longitudes during the 23rd solar cycle, a result that is coherent with the non-rejection of rotational symmetry for such cycle reported in \cite{rotasym}.\\

% T-PCA vs ST-PCA
An application of T-PCA, in either of its variants, proposed a mode of variation that is statistically less efficient (Figure \ref{fig:STPCA-TPCA_Variation} for SO ordering, analogous results for SI ordering). Furthermore, the deformation of $\mathbb{T}^3$ to $\mathbb{S}^3$ in T-PCA introduces distortion in the form of an artificial cluster structure on the first scores, as shown in Figures \ref{fig:sunspots_hist_TPCA_SI} and \ref{fig:sunspots_hist_TPCA_SO}, despite the marginals of $\{\boldsymbol\Theta_i\}_{i=1}^{5371}$ being fairly uniformly distributed. This contrasts with Figure \ref{fig:sunspots_hist_STPCA}. To provide a formal comparison between the distributions of the first scores of T-PCA and ST-PCA, we performed three omnibus circular uniformity tests using their asymptotic distributions as implemented in the \texttt{sphunif} R package \citep{sphunif}. The results are summarized in Table \ref{table:uniformity}. The resulting $p$-values illustrate that the uniformity hypothesis cannot be rejected for the marginal distribution of the original longitudes $\{\theta_i\}_{i=1}^{5373}$ and the corresponding scores for ST-PCA, for a significance level $\alpha = 0.05$, but is emphatically rejected with the scores obtained by the two possible orderings for~\mbox{T-PCA}.\\

% Final comparison
Finally, we provide a comprehensive comparison of ST-PCA with seven existing methods of dimensionality reduction on the torus for this dataset. To do so, we sampled a common subset of 1000 sunspots and computed for each method the percent of variance that is explained by each of its three components. Table \ref{table:var_explained} summarizes the results of that analysis, with several remarks following. First, dPCA gives a six-dimensional embedding and as a result, the sum of the variance explained by the first three dimensions does not add up to $100\%$. For dPCA+, histograms of ten bins were used to identify the regions of lowest density. Existing software for GeoPCA uses PGA and does not report the percentage of total variance explained, but the percentage of variance explained in the first two components instead; for that reason, we have not included it in the table. As a result, we were only able to find upper bounds for the percents of variance explained by the first two components ($73.54\%$ and $26.46\%$, respectively). Due to the apparent lack of appropriate software, the model-based approaches have not been implemented. As it can be seen, ST-PCA explains the most variance by the first component among its competitors. Its second component explains a smaller percentage compared to the third component, which is an often-encountered situation in non-Euclidean PCA (see Section \ref{subestction:var_explained}). For example, an instance of this issue for T-PCA is reported in Table 5(b) of \cite{Eltzner2018}. We conclude remarking that the methods that go through the sphere seem to yield a much higher signal compression compared to the other methods, with ST-PCA slightly outperforming T-PCA with regard to the percent of variance explained in the first component.

\begin{figure}[!h]
	\centering
	\subfloat[]{\includegraphics[width=0.32\textwidth,clip,trim={0cm 0cm 0cm 1.5cm}]{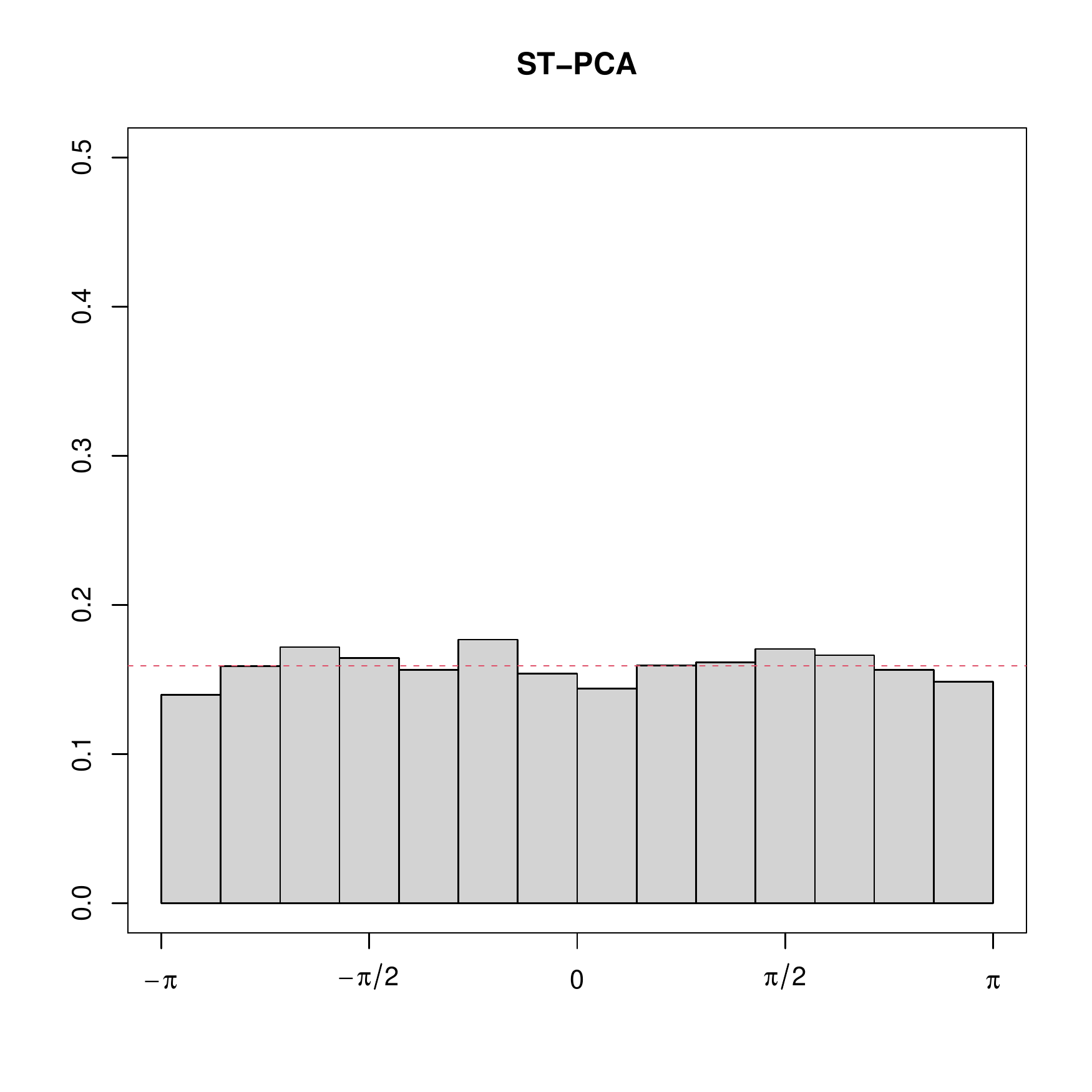} \label{fig:sunspots_hist_STPCA}}
	\hfill
	\subfloat[]{\includegraphics[width=0.32\textwidth,clip,trim={0cm 0cm 0cm 1.5cm}]{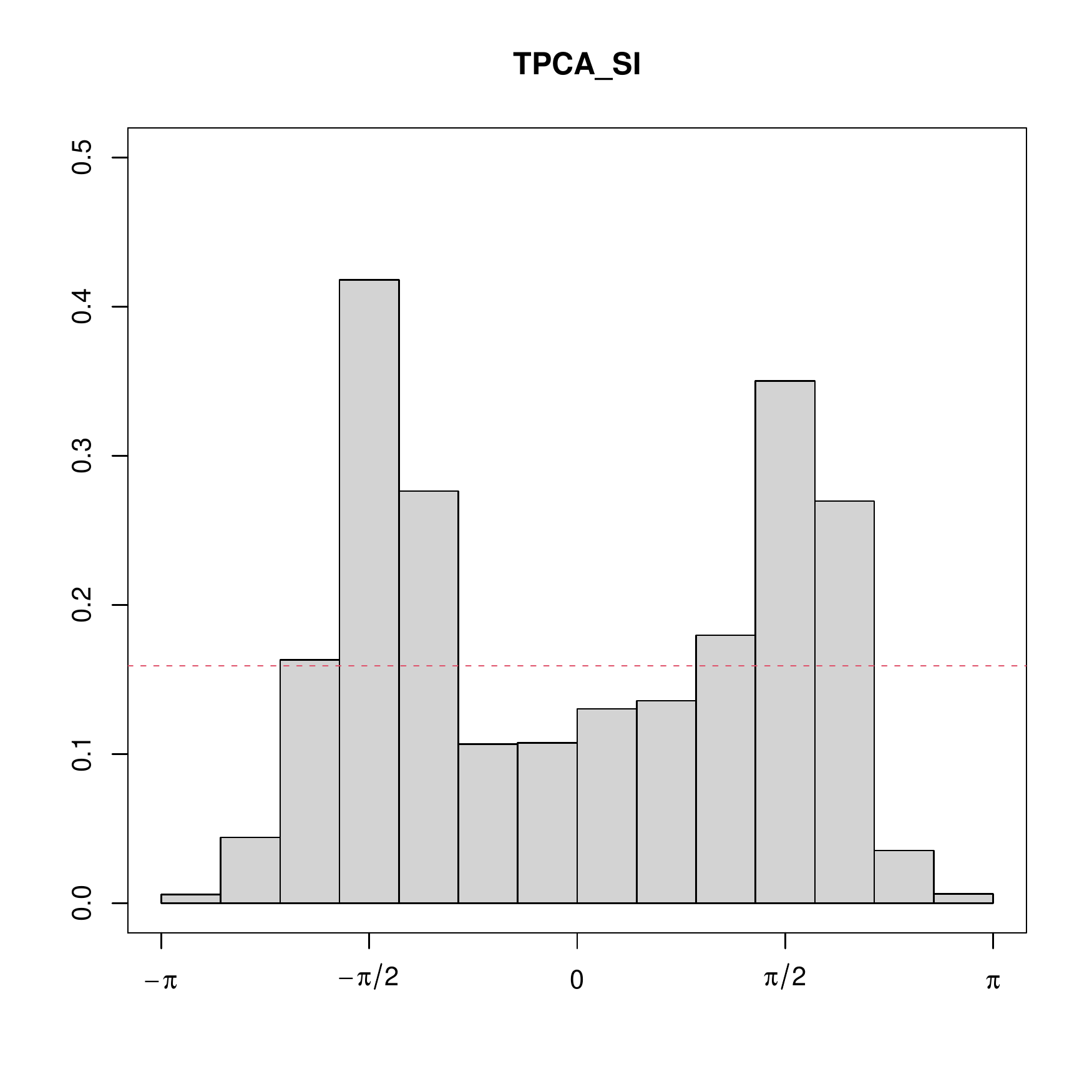}\label{fig:sunspots_hist_TPCA_SI}}
	\subfloat[]{\includegraphics[width=0.32\textwidth,clip,trim={0cm 0cm 0cm 1.5cm}]{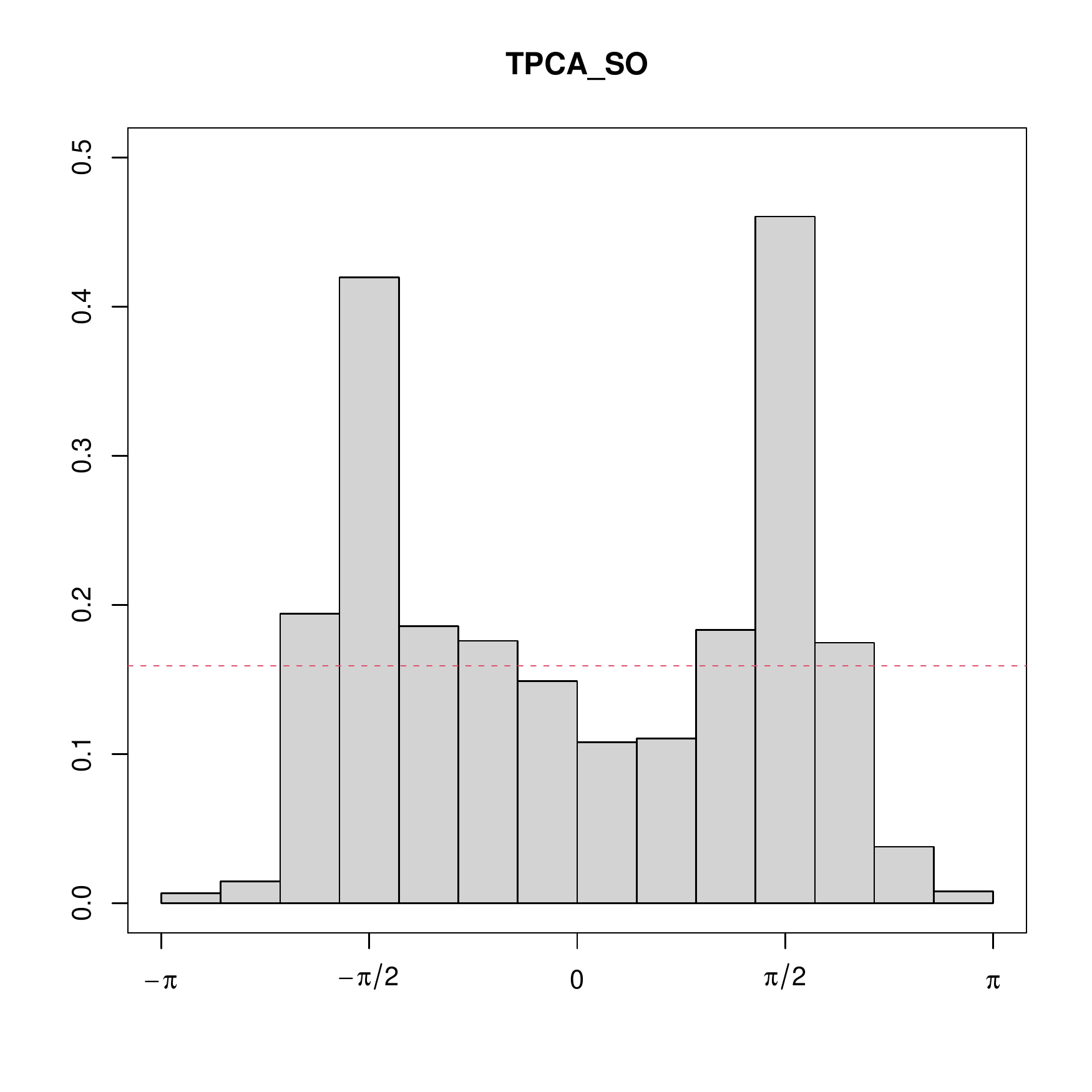}
		\label{fig:sunspots_hist_TPCA_SO}}
	\hfill
	\caption{\small Histograms of the first scores of ST-PCA and T-PCA: (a) ST-PCA; (b) T-PCA with SI ordering; (c) T-PCA with SO ordering. Both versions of T-PCA suggest the existence of spurious clusters (the marginals of the data are not significantly non-uniform), which are not present in ST-PCA.}
	\label{fig:sunspots_hist}
\end{figure}

\begin{table}[h!]
	\centering
	\small
	\begin{tabular}{|l|r|r|r|r|}
		\toprule
		Test & Longitudes & ST-PCA & T-PCA (SO) & T-PCA (SI) \\
		\midrule
		Giné's $F_n$ & 0.3215 & 0.0855 & 0 & 0\\
		Watson & 0.3595 & 0.1843 & 0 & 0\\
		Projected Anderson--Darling & 0.7554 & 0.1273 & 0 & $9.522\times10^{-9}$ \\
		\bottomrule
	\end{tabular}
	\caption{\small $p$-values for various circular uniformity tests. From left to right column: name of the test; $p$-values for testing uniformity on the original longitudes; $p$-values for testing uniformity on the first scores of the corresponding method. The original longitudes and ST-PCA's first scores are not significantly non-uniform, while uniformity is rejected in T-PCA's first scores.}
	\label{table:uniformity}
\end{table}

%---------------------------%
\subsection{An RNA dataset}
\label{subsec:RNA}
%---------------------------%

% RNA dataset description
This RNA dataset consists of nine angles, out of which six are dihedral $(\alpha,\beta,\gamma,\delta,\epsilon,\zeta)$, one corresponds to the base $(\chi)$, and two to the pseudotorsion angles $(\eta,\theta)$. This dataset was put together by \cite{Duarte1998} and updated by \cite{WADLEY2007942} in an endeavor to cluster the complex structure of RNA nucleotides. The pseudotorsion angles were found to approximate well their folding structure. \cite{Sargsyan2012} then curated a subset that consists of $190$ observations and three clusters (C3'-endo clusters I, II, and V of \cite{WADLEY2007942}, containing $59$, $88$, and $43$ points respectively), henceforth referred to as the \textit{small RNA dataset}. The $\eta$--$\theta$ plot of the pseudotorsion angles of the small RNA set reveals a clear cluster structure, which is not visible by the pairwise scatterplots of the other seven backbone and base angles (see, e.g., Figure 7 in \cite{Eltzner2018}). We want to investigate whether it is possible to predict the cluster structure revealed in the $\eta$--$\theta$ plot by using the remaining seven variables only. Following the analyses by \cite{Sargsyan2012}, \cite{Eltzner2018}, and \cite{nodehi2020torus}, the small RNA dataset has become a de facto benchmark for dimensionality reduction methods on the torus.

\begin{table}[!h]
	\centering
	\small
	\begin{tabular}{|l|r|r|r|}
		\toprule
		Method & Component 1 & Component 2& Component 3\\
		\midrule
		dPCA$\ssymbol{1}$ & 38.48 & 34.06 & 7.67 \\
		dPCA+$\ssymbol{1}$ & 42.69 & 30.53 & 26.78 \\
		aPCA$\ssymbol{1}$ & 43.27 & 31.28 & 25.45 \\
		Euclidean PCA$\ssymbol{1}$ & 55.43 & 24.48 & 20.09 \\
		Complex-dPCA$\ssymbol{1}$ & 72.51 & 14.59 & 12.90 \\
		T-PCA (SI)$\ssymbol{2}$ & 88.27 & 6.97 & 4.76 \\ 
		T-PCA (SO)$\ssymbol{2}$ & 89.32 & 6.43 & 4.25 \\
		ST-PCA$\ssymbol{2}$ & 90.51 & 3.07 & 6.40 \\
		\bottomrule
	\end{tabular}
	\caption{\small Percentage of variance explained by components of dimension reduction methods on the torus, sorted decreasingly according to the percentage of variance explained by their first components. ST-PCA and T-PCA explain the most percentage of variance in their first component. $\ssymbol{1}$Variance calculations based on standard PCA. $\ssymbol{2}$Variance calculations based on torus variance as defined in Section \ref{subestction:var_explained}.}
	\label{table:var_explained}
\end{table}

% Previous analyses
Figures 3 and 6 of \cite{Sargsyan2012} reveal that dPCA fails at accomplishing any separation by utilizing the dihedral and the base angles, while GeoPCA succeeds in separating cluster I from II in the first two principal components. However, as it can be seen from the $\eta$--$\theta$ plot, cluster V is close to cluster II and GeoPCA can offer no further insight on such a separation based on the first two components. An analysis by TPPCA on a subset of the small RNA dataset consisting of 181 nucleotides can give a decent separation using the first two components. Once more, however, the distinction between the two nearby clusters is not very clear (see Figure 2 of \cite{nodehi2020torus}). Out of torus dimensionality reduction techniques, T-PCA on a 181 nucleotides subset, followed by mode hunting, has been the most successful at revealing the cluster structure of all 3 clusters by using only the first principal component (see Figure 8 of \cite{Eltzner2018}).\\

% Comparison with T-PCA I
To produce a clear comparison of ST-PCA's performance to T-PCA, we obtain the spherical scores of the $190$ nucleotides dataset with ST-PCA ($\widehat{r}=1.49$), as well as T-PCA with SI and SO ordering (version ``20200518'' as provided by the authors). Then, we proceed with a modified clustering scheme (to accommodate for the periodicity of the data) based on kernel density estimation, on the scores of the first principal components. The bandwidth has been determined using the plug-in selector for density derivative estimation \citep[pages 49 and 70]{Wand1995} implemented in the R package \texttt{ks} \citep{ks}. Figure \ref{fig:RNA} demonstrates the results of our analysis, where the colors of the points correspond to their true labels: red, yellow, and green are used respectively for clusters I, II, and V. The domains of attraction of each cluster are represented by a different color in the scores plots and are separated by dashed vertical lines.\\

% Comparison with T-PCA II
The left column of Figure \ref{fig:RNA} demonstrates that deforming the torus with the SI scheme underperforms by recognizing two clusters, as well as three ill-defined. In particular, it fails to separate clusters II and V and thus misclassifies 105 points (classification rate: $0.447$). The middle column shows that T-PCA with SO ordering correctly recognizes all three clusters, while misclassifying 23 points (classification rate: $0.879$). Finally, the right column shows that ST-PCA also reveals three clusters and misclassifies $16$ points (classification rate: $0.916$), hence outperforming T-PCA in both its variants. An additional benefit for ST-PCA is that the distribution of the estimated densities suggests a more clear separation of the clusters.

\begin{figure}[!h]
	\vspace*{-0.5cm}
	\centering
	\includegraphics[width=\textwidth]{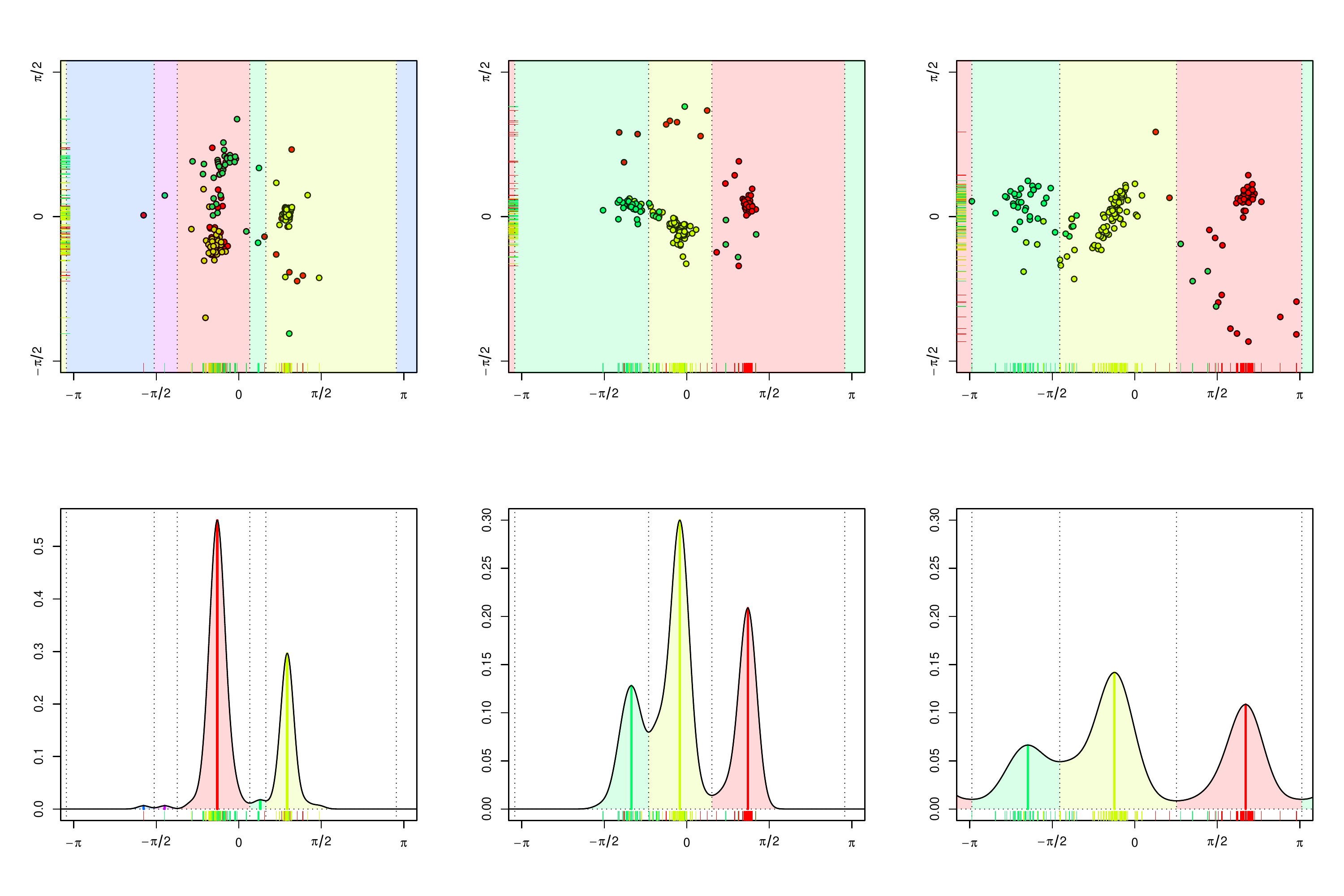}
	\caption{\small Upper row: scores scatterplots for the first two components; first component on the horizontal axis, second on the vertical. Lower row: kernel density estimation investigation of modality for the first scores. From left to right: T-PCA with SI ordering; T-PCA with SO ordering; ST-PCA. Shows that ST-PCA gives the best separation of these clusters in the first scores.}
	\label{fig:RNA}
\end{figure}

%---------------------------%
\section{Discussion}
\label{sec:epilogue}
%---------------------------%

In this paper, a novel approach to dimensionality reduction on the torus has been developed. It is an analogue of PCA, as it provides modes of variations and scores. Consistent with previous work, we have provided evidence that for toroidal data a transformation to the sphere induces less distortion compared to direct Euclidean transformations. We propose a fully data-driven transformation that is optimal in terms of minimizing the stress function inherited from the MDS literature. This key advantage of ST-PCA comes also at a price: the inheritance of the computational limitations of MDS for larger datasets and the delicate inversion procedure of such transformation. ST-PCA's benefits are validated by real data applications on the appearance of sunspots and an benchmark RNA dataset, where ST-PCA is seen to overperform previous methods in both qualitative and quantitative metrics.

%---------------------------%
\section*{Acknowledgments}
%---------------------------%

We are grateful to Benjamin Eltzner for kindly sharing his implementation of T-PCA. The second author acknowledges support by grants PGC2018-097284-B-100 and IJCI-2017-32005 by Spain's Ministry of Science, Innovation and Universities. Both grants are co-funded with FEDER~funds.

%\bibliographystyle{apalike}
%\bibliography{Pavlos.bib}

\end{document}